\documentclass[12pt,notoc]{JHEP}

\usepackage{amsmath,amssymb,euscript,array,cite,mathrsfs}

\usepackage{epsfig}

\def\XXint#1#2#3{{\setbox0=\hbox{$#1{#2#3}{\int}$}
     \vcenter{\hbox{$#2#3$}}\kern-.5\wd0}}

\newcommand{\IM}{\operatorname{Im}}

\def\a{\alpha}
\def\b{\beta}
\def\c{\gamma}
\def\d{\delta}
\def\e{\epsilon}

\def\l{\lambda}
\def\m{\mu}
\def\n{\nu}

\def\r{\rho}
\def\s{\sigma}
\def\t{\tau}

\def\w{\omega}

\def\D{\Delta}
\def\L{\Lambda}

\def\Dbarslash{\,\,{\raise.15ex\hbox{/}\mkern-12mu {\bar D}}}
\def\Dslash{\,\,{\raise.15ex\hbox{/}\mkern-12mu D}}
\def\delslash{\,\,{\raise.15ex\hbox{/}\mkern-9mu \partial}}
\def\delbarslash{\,\,{\raise.15ex\hbox{/}\mkern-9mu {\bar\partial}}}

\def\rta{\rightarrow}

\title{{\Large Polarized Structure Functions and Two-Photon Physics at
    Super-B }}
\author{G. M. Shore\\
Department of Physics,\\ Swansea University,\\
Swansea, SA2 8PP, UK.\\
E-mail: {\tt g.m.shore@swansea.ac.uk}}
\abstract{
The potential of polarized, high-luminosity, moderate-energy 
$e^+ e^-$ colliders for performing unique measurements in 
fundamental QCD is described, with particular reference to the
proposed Super-B facility.  An extensive programme of 2-photon 
physics is proposed, focusing on measurements of the polarized 
photon structure functions $g_1^\gamma$ and $g_2^\gamma$ and
pseudoscalar meson transition functions. The experimental 
requirements for Super-B to make the first measurement of the first
moment sum rule for the off-shell polarized photon structure function 
$g_1^\gamma(x,Q^2;K^2)$ are described in detail. Cross-section
formulae and experimental issues for investigations of NLO and higher-twist
effects in $g_1^\gamma$ and $g_2^\gamma$ together with 
exclusive 2-photon meson production are presented. 
This programme of QCD studies complements the core mission of 
Super-B as a high-luminosity B factory investigating flavour physics
and rare processes signaling new physics beyond the standard model. 

}

\begin{document}

\section{Introduction}

High-luminosity, moderate energy $e^+ e^-$ colliders open a window on
a wide and interesting range of phenomena in QCD.  They provide an
especially clean environment where many fundamental aspects of QCD
itself can be studied without the complication of bound-state hadronic
targets. With polarized beams, they provide the ideal conditions to
study the polarized photon structure functions $g_1^\gamma$ and
$g_2^\gamma$, including moment sum rules and higher-order perturbative
QCD and higher-twist effects, $U(1)_A$ dynamics and anomalies, the
gluon topological susceptibility, exclusive pseudoscalar meson
production and transition functions, chiral symmetry breaking and
vector meson dominance, amongst many others.

Super-B \cite{Bona:2007qt,O'Leary:2010af,Biagini:2010cc,Grauges:2010}
is a high-luminosity, asymmetric $e^+ e^-$ collider to be
built at the Cabibbo Laboratory at the University of Rome 
`Tor Vergata' campus, with commissioning
expected in 2017. It is conceived as a B factory, with asymmetric
$e^+$ and $e^-$ beams with CM energy initially tuned to the
$\Upsilon(4S)$ resonance at $\sqrt{s}= 10.58~ {\rm GeV}$ and luminosity
$10^{36}~{\rm cm}^{-2} {\rm s}^{-1}$ corresponding to an annual
integrated luminosity in excess of $12~{\rm ab}^{-1}$. It is designed
to study precision flavour physics and rare events with a view to 
discovering signals of new physics beyond the standard model. 
The extensive scope of this physics programme is described in detail 
in ref.\cite{O'Leary:2010af} while descriptions of the accelerator
and detector can be found in refs.\cite{Biagini:2010cc} and 
\cite{Grauges:2010} respectively. Importantly, it is
planned that from the outset the electrons in the low-energy ring
will be polarized, with efficiencies of over 70\% \cite{Wienands:2010zz}.
This will be achieved by injecting tranversely polarized electrons 
and using a system of spin rotators to produce a longitudinally 
polarized beam within the interaction region.

In this paper, we point out that in addition to its core flavour
physics mission, Super-B has the potential to perform unique studies
of polarization phenomena in QCD through a complementary programme
of polarized 2-photon physics.  To illustrate this potential, we
describe in detail a number of QCD measurements, focusing on the
polarized photon structure functions $g_1^\gamma$ and $g_2^\gamma$ and
pseudoscalar meson transition functions, explaining the accelerator
and detector requirements for these to be made at Super-B.

Foremost amongst these is the first moment sum rule for the polarized
photon structure function $g_1^\gamma(x,Q^2;K^2)$, where $Q^2$ and
$K^2$ are the invariant momenta of the scattered and target photon
respectively, as measured in the inclusive process 
$e^+ e^- \rightarrow e^+ e^- X$  in the deep-inelastic
regime (see Fig.~1).  This was first proposed by
Narison, Shore and Veneziano in 1992 \cite{Narison:1992fd,Shore:1992pm},
though only now has collider
technology evolved to the point where a detailed experimental
verification has become possible.  
We emphasise that, in contrast to experiments using real
back-scattered laser photons as the target, the target photons in 
$e^+ e^-$ scattering are in principle virtual and indeed almost all the
interesting QCD physics resides in the $K^2$-dependence of the sum
rule. The sum rule can be written as \cite{Narison:1992fd,Shore:2004cb}:
\begin{multline}
\int_0^1 dx~g_1^\c(x,Q^2;K^2) = {1\over18}{\a\over\pi}
\left(1-{\a_s(Q^2)\over\pi}\right)  \\
\times \biggl[3 F^3(K^2) + F^8(K^2) + 8 F^0(K^2;\m^2) 
\exp\int_{t(K^2)}^{t(Q^2)} dt' \c(\a_s(t'))\biggr] \ ,
\label{aa}
\end{multline}
in terms of non-perturbative form factors $F^a(K^2)$ which
characterise the anomalous three-current AVV correlation function
$\langle 0|J_{\m 5}^a(0)~J_\l(k)~J_\r(-k)|0\rangle$.
Here, $\c(\a_s)$ is the  anomalous dimension of the flavour singlet
axial current, $t(Q^2) = {1\over2}\log(Q^2/\mu^2)$, 
and we assume three dynamical quark flavours, so 
$a = 3,8$ denote $SU(3)$ flavour generators with $a=0$ the singlet.
The AVV correlator is an important quantity in non-perturbative QCD
and encodes a wealth of information about anomalies, chiral symmetry
breaking and the validity of widely used models such as vector meson
dominance.  First-principles theoretical calculations are challenging 
and the opportunity to compare with direct experimental measurements 
for a variety of external momenta will be valuable.

For real photons, $K^2=0$, electromagnetic gauge invariance implies
the simple sum rule
\begin{equation}
\int_0^1 dx~ g_1^\c(x,Q^2;0) = 0 \ ,
\label{ab}
\end{equation}
first derived by Bass \cite{Bass:1991sg}
 (see also refs.\cite{Narison:1992fd, Shore:1992pm, Bass:1998bw}). 
For target photons with invariant momenta in the range $m_\rho^2 \ll K^2
\ll Q^2$, the sum rule is determined entirely by the electromagnetic
$U(1)_A$ anomaly with perturbative QCD corrections given by Wilson
coefficients together with the anomalous dimension related to the
QCD $U(1)_A$ anomaly. It was shown in ref.\cite{Narison:1992fd} 
that to NLO, {\it i.e.}~$O(\a\a_s)$,
\begin{equation}
\int_0^1 dx~ g_1^\c(x,Q^2;K^2) = {2\over3}{\a\over\pi}
\biggl[1 - {\a_s(Q^2)\over\pi} + {4\over9}\biggl(
 {\a_s(Q^2)\over\pi} -  {\a_s(K^2)\over\pi}\biggr)\biggr] \ .
\label{ac}
\end{equation}
Note that the overall normalisation factor is $N_c \sum_f \hat e_f^4$,
proportional to the fourth power of the quark charges $\hat e_f$,
corresponding to the lowest order box diagram contributing to
$g_1^\c$. This result was verified in
refs.\cite{Sasaki:1998vb,Ueda:2006cp} 
and subsequently
extended to NNLO, $O(\a\a_s^2)$, in ref.\cite{Sasaki:2006bt}.

In order to verify the first moment sum rule experimentally, we
require polarized beams and a sufficiently high luminosity to allow
the spin asymmetry of the cross-section to be measured, recalling 
\cite{Narison:1992fd} that it is kinematically suppressed by a factor of 
$Q_{\rm min}^2/s$ relative to the total cross-section.  This factor
also explains why colliders with moderate CM energy $\sqrt{s}$ are
favoured for this type of QCD spin physics. Identification of the
target photon virtuality $K^2$ is most clearly done by tagging the
target electron,\footnote{For simplicity, we use the term `electron'
to denote either the electron or positron beam.} though this is 
experimentally challenging for the small angles necessary to access
the non-perturbative region $K^2\simeq m_\rho^2$. The perturbative 
sum rule \eqref{ac}, for $K^2 \gtrsim 1~{\rm GeV}^2$ is more 
readily measurable. These experimental issues are discussed in
detail in section 5, after we derive the relevant cross-section moment
formulae in section 2.  It is important here that all these formulae
are derived without use of the conventional `equivalent photon'  
formalism (see, {\it e.g.}~refs.~\cite{Berger:1986ii,Brodsky:1971ud}, 
since the $K^2$-dependence of the target photon is crucial.

If the azimuthal angle between the planes of the scattered and target
electrons is also measured, then we can identify the second polarized
photon structure function $g_2^\c(x,Q^2;K^2)$. This is of additional
theoretical interest since it receives contributions from both twist 2
and twist 3 operators in the OPE analysis of deep inelastic
scattering. Following ref.\cite{Baba:2002mx}, we show how to isolate the
twist 3 contribution, then derive cross-section formulae to use the
azimuthal angle dependence of the scattered electrons to distinguish
the structure functions and determine $g_2^\c$.

In refs.\cite{Narison:1992fd,Shore:1992pm,Shore:2004cb,Shore:2007yn}, 
it was shown how the non-perturbative form
factors $F^a(K^2)$ can be related to the off-shell transition
functions $g_{P\c^*\c^*}(0,K^2,K^2)$ of the pseudoscalar mesons
$P=\pi, \eta, \eta'$.  An important subtlety arises in the flavour
singlet sector, where the QCD $U(1)_A$ anomaly means that the
equivalent result for the $\eta'$ also involves the gluon topological
susceptibility, the key non-perturbative quantity which controls much
of the $U(1)_A$ dynamics of QCD. 
In fact, the pseudoscalar meson transition functions can be measured
directly for different photon virtualities via the exclusive
two-photon production reaction $e^+ e^- \rightarrow e^+ e^- P$
(see Fig.~2) even with unpolarized beams. Such measurements have
already been made at CELLO \cite{Behrend:1990sr}, CLEO
\cite{Gronberg:1997fj}
and BABAR \cite{Aubert:2009mc,BABAR:2011ad,Lees:2010de}
for transition functions $g_{P\c^*\c}(m_P^2,Q^2,0)$ with one virtual 
and one assumed real photon. Here, in section 4, we derive
cross-section formulae relevant to polarized beams and discuss what 
may be learned more generally from measurements of meson transition 
functions at Super-B. In addition to their intrinsic interest,
these are important in theoretically determining 
the virtual light-by-light $\c^* \c^* \rightarrow \c^* \c^*$
scattering amplitude, which is itself a key part of the hadronic
contribution which is the major uncertainty in reconciling theoretical 
predictions with experimental measurements of $g-2$ for the muon
\cite{Jegerlehner:2009ry}.

This theoretical analysis of the polarized photon structure functions
and pseudoscalar meson transition functions is presented in sections
2-4, with extensive reference to our earlier papers
\cite{Narison:1992fd,Shore:1992pm,Shore:2004cb,Shore:2007yn,
Shore:1991np,Shore:1999tw,Shore:2006mm}.
See also refs.~\cite{Sasaki:1998kb,Stratmann:1999bv,Sasaki:1999py,
Sasaki:2000tm,Kwiecinski:2000yk,Gluck:2001az,Baba:2003ed,
Baba:2002mx,Ueda:2006cp,Sasaki:2006bt,Watanabe:2011xb}
for a selection of further papers on the $g_1^\c$ and $g_2^\c$ photon
structure functions, mainly from a parton perspective.
Here, our focus
is on deriving cross-section formulae and investigating the
experimental requirements to measure $g_1^\c(x,Q^2;K^2)$,
$g_2(x,Q^2;K^2)$ and $g_{P\c*\c*}$ at a high-luminosity, polarized 
$e^+ e^-$ collider. In the final section, we turn more specifically to 
Super-B and investigate the cross-sections and experimental cuts 
necessary to measure the first moment sum rule for $g_1^\c$ with 
~the design CM energy and luminosity.  We will also consider what 
detector requirements are necessary to realise the full potential of 
Super-B as the collider of choice to investigate polarized QCD 
phenomenology.

\section{Polarized Photon Structure Functions $g_1^\gamma$ and
  $g_2^\gamma$ }

In this section, we show how to determine the polarized photon
structure functions $g_1^\c(x,Q^2;K^2)$ and $g_2^\c(x,Q^2;K^2)$ 
and their moments from the inclusive process 
$e^+ e^- \rightarrow e^+ e^- X ~{\rm (hadrons)}$
shown in Fig.~1.

\begin{figure}[ht]
\centerline{\includegraphics[width=2.3in]{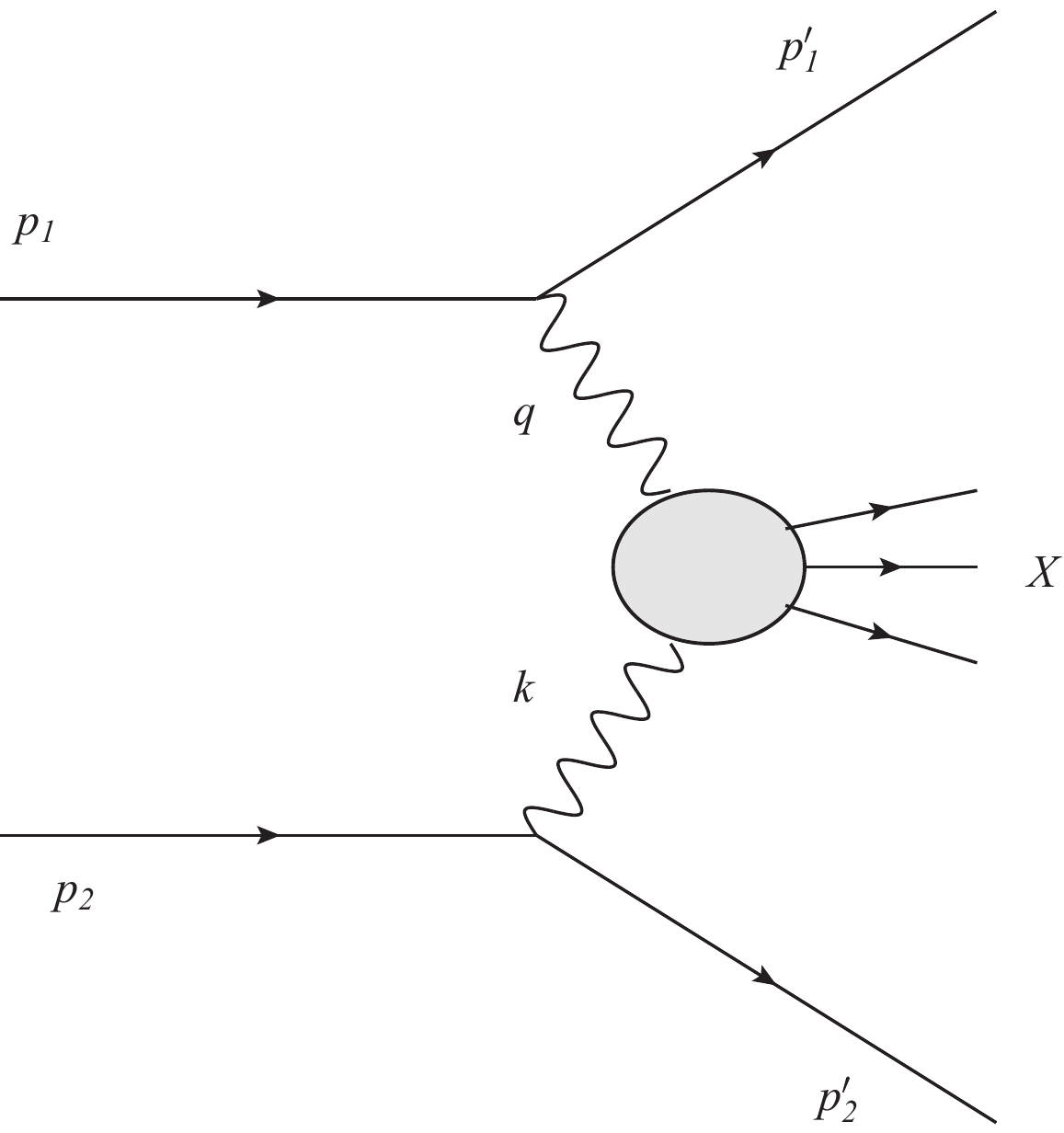}}
\caption{\footnotesize Kinematics for the inclusive two-photon
  reaction $e^+ e^- \rta e^+ e^- X ~{\rm hadrons}$.}
\label{Fig1}
\end{figure}

We begin with some kinematics. The total cross-section is given by
integrating over the phase space of the scattered electrons, a total
of 6 degrees of freedom.  However, by symmetry only the relative
azimuthal angle $\phi = \phi_1 - \phi_2$ of the scattering planes of
the electrons is physically relevant, so we need to specify only 5
Lorentz invariants. We choose these to be $Q^2 = -q^2$, $K^2 = -k^2$,
$\nu_e = p_2.q$, $\bar\nu_e = p_1.k$ and $\nu = k.q$. The
deep-inelastic limit is $Q^2,\nu_e,\nu \rightarrow \infty$ with
$x_e = Q^2/2\nu_e$ and $x = Q^2/2\nu$ fixed. Note that while we use 
the analogous notation for the `target' electron, neither $K^2$ nor
$\bar\nu_e$ is assumed to be large, and indeed we are most interested 
in the regime $K^2/Q^2\ll 1$, with $K^2$ less than around 1 GeV$^2$.

In terms of the momenta and angles of the scattered electrons in the 
lab frame, we define
$p_1^\mu = (E_1, 0, 0, E_1)$ and $p_1^{\prime\mu} = (E_1', E_1'\sin\theta_1
\cos\phi_1, E_1'\sin\theta_1 \sin\phi_1, E_1'\cos\theta_1)$, 
in the limit where we neglect the electron mass, and
similarly for $p_2^\mu$ and $p_2^{\prime\mu}$, noting that for an
asymmetric collider like Super B, $E_1\neq E_2$.
We then have, with $s = (p_1+p_2)^2 = 4E_1E_2$, 
\begin{align}
Q^2 & = 4 E_1 E_1' \sin^2{\theta_1\over2}~~~~~~~~~~~~~~~~
~~~~~~~~~
K^2 = 4 E_2 E_2' \sin^2{\theta_2\over2} \nonumber \\
\nu_e & = 2E_2\left( E_1 - E_1' \cos^2{\theta_1\over2}\right) ~~~~~~~~~~~~~~
\bar\nu_e  = 2 E_1 \left(E_2 -  E_2' \cos^2{\theta_2\over2}\right) \nonumber \\
\nu & = \frac{2}{s} \bigl(\nu_e \bar\nu_e + {1\over4}K^2 Q^2\bigr) + k_T q_T
\cos\phi  
\label{ba}
\end{align}
where we define $q_T (k_T)$ as the magnitude of the transverse
component of $q^\mu (k^\mu)$, so that
\begin{equation}
q_T^2 = Q^2 \Bigl(1 - {2\nu_e\over s}\Bigr) \ ,~~~~~~~~~~~~~~~~~~
k_T^2 = K^2 \Bigl(1 - {2\bar\nu_e\over s}\Bigr) \ .
\label{bb}
\end{equation}
The on-shell condition for the electrons determines $p_1.q = -
{1\over2}Q^2$ and $p_2.k = -{1\over2}K^2$ so these are not independent
variables. Also note that $\n = \tfrac{1}{2}(Q^2 + K^2 + W^2)$, where 
$W^2$ is the total hadronic invariant momentum.

The complete set of 5 kinematic variables is therefore the measurable
quantities $E_1', \theta_1, E_2', \theta_2,\phi$, which
determine the invariants $Q^2, K^2, \nu_e,\bar\nu_e,\nu$. 
We will be concerned with cross-sections that are differential with 
respect to different subsets of these 5 invariants. Note that the 
azimuthal dependence $\phi$ is encoded in $\bar\nu_e$ once $\nu_e$
and $\nu$ are fixed. For the theoretical discussion, we assume that
the target electron is tagged, with $E_2'$ and $\theta_2$ (as well as
$\phi$) being measured directly. The experimental issue of whether
this is possible for the interesting range of $K^2$, or whether it 
would be necessary to try to reconstruct the
scattered target electron momentum from a knowledge of  
the total hadronic momentum, is discussed later.

The photon structure functions are defined from the Green function
$T_{\mu\nu\l\r}(q,k) = \langle 0|J_\mu(q) J_\nu(-q) A_\l(k)
A_\r(-k)|0\rangle$.\footnote{Note that we use the Green function
  with $A_\l$, $A_\r$ rather than four currents to allow for the
  direct coupling from the operators involving $F^2$ and $F\tilde F$ 
  occurring in the OPE for $J_\mu(q) J_\nu(-q)$.}
There are two independent Lorentz structures,
which we denote \cite{Sasaki:1998vb,Ueda:2006cp} 
as $I_{\mu\nu\l\r}$ and $J_{\m\nu\l\r}$ where
\begin{align}
I_{\mu\nu\l\r} & = - \nu \epsilon_{\mu\nu\a\c}\epsilon_{\l\r\b}{}^\c
q^\a k^\b  \nonumber \\
J_{\mu\nu\l\r} & = I_{\mu\nu\l\r} -
\epsilon_{\mu\nu\a\c}\epsilon_{\l\r\b\d} q^\a q^\b k^\c k^\d  \ .
\label{bc}
\end{align}
The structure functions $g_1^\c$ and $g_2^\c$ are then given in terms
of the antisymmetric part of $T_{\mu\nu\l\r}$ as follows:
\begin{equation}
\tfrac{1}{\pi} \IM T_{\mu\nu\l\r}^A = {1\over {2K^4 \nu^2}} 
\Bigl[ g_1^\c(x,Q^2;K^2)~ I_{\mu\nu\l\r}  ~+~  g_2^\c(x,Q^2;K^2)~ 
J_{\mu\nu\l\r}\Bigr] \ .
\label{bd}
\end{equation}

The usual leptonic tensor for electrons with helicity 
$h_1/2$, where $h_1 = \pm 1$, is defined as:
\begin{equation}
L_{\l\r}(p_1,h_1;q) = 4 p_{1\l} p_{1\r} - 
2(p_{1\l} q_\r - p_{1\r}q_\l) + q^2 g_{\l\r}
+ 2i h_1 \epsilon_{\l\r\s\t}p_1^\s q^\t   \ ,
\label{be}
\end{equation}
and it is convenient to define the antisymmetric part,
\begin{equation}
L^A_{\l\r}(p_1;q) = 2i \epsilon_{\l\r\s\t}p_1^\s q^\t \ .
\label{bee}
\end{equation}
We then find
\begin{equation}
\tfrac{1}{\pi}\IM T_{\mu\nu\l\r}^A(q,k) L^{\l\r}(p_2,h_2;k) =
- 2i h_2  K^2 {1\over\nu}\epsilon_{\mu\nu\a\c}q^\c
\left[g_1^\c (p_2^\a -\tfrac{1}{2}k^\a) 
+ g_2^\c \left(p_2^\a - {\nu_e\over\nu}k^\a\right) \right]
\label{bf}
\end{equation}
and (compare ref.\cite{Baba:2002mx}, appendix A)
\begin{multline}
L^{\mu\nu}(p_1,h_1;q) \tfrac{1}{\pi} \IM T_{\mu\nu\l\r}^A(q,k)
L^{\l\r}(p_2,h_2;k) \\
= - 4 h_1 h_2 K^2 Q^2 \frac{1}{\nu^2}
\left[g_1^\c (s + \tfrac{1}{2}\nu - \nu_e - \bar\nu_e)
+ g_2^\c \left(s - \frac{2\nu_e \bar\nu_e}{\nu}\right) \right] \ .
\label{bg}
\end{multline}

The cross-section polarization asymmetry 
$\D\s = {1\over2}\bigl(\s(++) - \s(+-)\bigr)$
is then given as an integral over $q$ and $k$ as follows:
\begin{multline}
\D\s = \frac{\a^3}{4\pi^2}  \frac{1}{s} \int d^4 q 
~\d\left(p_1.q + \tfrac{1}{2}Q^2\right) ~\int d^4 k~
\d\left(p_2.k + \tfrac{1}{2}K^2\right)   \\
\times 
~\frac{1}{Q^4} L_A^{\mu\nu}(p_1;q) \tfrac{1}{\pi}
\IM T_{\mu\nu\l\r}^A(q,k)
L_A^{\l\r}(p_2;k) \ .
\label{bh}
\end{multline}
Changing variables in the $\int d^4 q$ integration, including the
Jacobian factor $1/s$, and integrating over the irrelevant azimuthal
angle $\int d\phi_1 = 2\pi$ using the cylindrical symmetry, we find
\begin{multline}
\D\s = {\a^3\over2\pi} ~ \frac{1}{s^2} \int dQ^2 \int d\nu_e
\int d^4 k ~\d\left(p_2.k + \tfrac{1}{2}K^2\right)  \\
\times \frac{1}{Q^4}
 L_A^{\mu\nu}(p_1;q) \tfrac{1}{\pi}\IM T_{\mu\nu\l\r}^A(q,k)L_A^{\l\r}(p_2;k) \ .
\label{bi}
\end{multline}

\vskip0.3cm
\noindent{\bf (i)~~$g_1^\c(x,Q^2;K^2)$ and cross-section moments:}

The first structure function $g_1^\c(x,Q^2;K^2)$ and its moments can be isolated
by measuring the differential cross-section $d^3\D\s / dQ^2 dx_e dK^2$.
To see this, note that under the full $\int d^4 k$ integral
we can effectively substitute $\bar\nu_e \rightarrow s\nu/2\nu_e
+ O(K^2/Q^2)$. The $g_2^\c$ dependence then drops out of 
eq.\eqref{bg},\footnote{In general, we may write
\begin{equation*}
\int d^4 k~ g(x,Q^2;K^2) ~k^\a = A p_2^\a + B q^\a
\end{equation*}
where, by making appropriate contractions,
\begin{equation*}
A =  \int d^4 k ~g(x,Q^2;K^2) ~{\nu\over\nu_e} \left(1 - {1\over2}
{K^2 Q^2\over\nu\nu_e}\right),~~~~~~~~~~~~
B = \int d^4 k ~g(x,Q^2;K^2) \left(-{1\over2}{K^2\over\nu_e}\right) \ .
\end{equation*}
It follows immediately that 
\begin{equation*}
\int d^4 k ~g(x,Q^2;K^2) \bar\nu_e = \int d^4 k ~g(x,Q^2;K^2)~
{s\nu\over2\nu_e} \left(1 - {1\over2}{K^2 Q^2\over\nu\nu_e}
+ {1\over2}{K^2 Q^2\over s\nu}\right) \ .
\end{equation*}
} while the coefficient of $g_1^\c$ reorganises into the
familiar Altarelli-Parisi splitting function $\D P_{\c e}(z) = 2 -z$,
where $z = x_e/x$, and the standard kinematical factor
$\left(1 - Q^2/2x_e s\right)$ which arises in polarized
cross-sections.  Then, rewriting the $\int d^4k$ in terms of the
invariants $K^2, \nu$ and the azimuthal angle $\phi_2$, which we can
then integrate over, we finally find:
\begin{equation}
\D\s = {\a^3\over s} \int_0^\infty{dQ^2\over Q^2} \int_0^1{dx_e\over
  x_e}  \int_0^\infty{dK^2\over K^2} \int_{x_e}^1 {dx\over x} ~
\D P_{\c e}\left({x_e\over x}\right) ~g_1^\c(x,Q^2;K^2) ~
\left(1 - {Q^2\over 2x_e s}\right) \ ,
\label{bj}
\end{equation}
reproducing the result quoted in 
refs.\cite{Narison:1992fd,Shore:2004cb}.\footnote{
Alternatively, as in refs.\cite{Narison:1992fd,Shore:2004cb}, 
we may apply the result in footnote 3 
at the point of defining the electron analogue of the usual hadronic
tensor in deep-inelastic scattering, together with `electron structure
functions' $g_1^e (x_e,Q^2)$, as follows:
\begin{equation*}
W_{\mu\nu}^A = {\a\over(8\pi)^2} \int d^4 k ~\d\bigl(p_2.k +
{1\over2}K^2\bigr) L^{\l\r}(p_2,+;k) \tfrac{1}{\pi}\IM T_{\mu\nu\l\r}^A(q,k) = 
{i\over4\nu_e} \epsilon_{\mu\nu\a\c}p_2^\a q^\c g_1^e(x_e,Q^2) \ ,
\end{equation*}
where 
\begin{equation*}
g_1^e(x_e,Q^2) = {\a\over2\pi} \int_0^\infty{dK^2\over K^2}
\int_{x_e}^1 {dx\over x} \D P_{\c e}\left({x_e\over x}\right) 
g_1^\c(x,Q^2;K^2) \ ,
\end{equation*}
and we again see that only the dependence on $g_1^\c$ survives.}
In particular, 
$x$-moments of the structure function $g_1^\c(x,Q^2;K^2)$ can be
measured in terms of the $x_e$-moments of the polarization asymmetry
of the differential cross-section as follows:
\begin{equation}
\int_0^1 dx_e~x_e^n~{d^3\D\s\over dQ^2 dx_e dK^2} =
{\a^3\over s Q^2 K^2} \int_0^1 dz~z^{n-1} \D P_{\c e}(z)~
\int_0^1 dx~ x^{n-1}~g_1^\c(x,Q^2;K^2) \ ,
\label{bk}
\end{equation}
where the integral over the splitting function factorises. 
(For simplicity, we have assumed here that $Q^2/2x_e s \ll 1$
though this factor could easily be retained.)
In particular, for the first moment we have simply
\begin{equation}
\int_0^1 dx_e~{d^3\D\s\over dQ^2 dx_e dK^2} =
{3\over2}{\a^3\over s Q^2 K^2} 
\int_0^1 dx~g_1^\c(x,Q^2;K^2) \ ,
\label{bl}
\end{equation}
the key point being that the cross-section is differential w.r.t.~the
standard DIS variables $Q^2,x_e$ and the target photon virtuality
$K^2$ only, but with the dependence on the azimuthal angle 
$\phi$ integrated out.

\vskip0.3cm
\noindent{\bf (ii)~~Azimuthal dependence and $g_2^\c(x,Q^2;K^2)$:}

Alternatively, we may retain the explicit dependence on the azimuthal
scattering angle in the differential cross-sections, which allows us
to measure the second polarized structure function
$g_2^\c(x,Q^2;K^2)$.  This time, we re-express the $\int d^4 k$ in 
eq.(\ref{bi}) directly as an integral over the invariants $K^2,\nu$
and $\bar\nu_e$, which encodes the $\phi$-dependence. 
Rearranging terms, we find:
\begin{multline}
\D\s = \frac{\a^3}{\pi} \frac{1}{s} 
\int_0^\infty \frac{dQ^2}{Q^2} \int_0^1 \frac{d x_e}{x_e} 
\int_0^\infty \frac{dK^2}{K^2} \int_0^1 \frac{d\bar x_e}{\bar x_e} 
\int_{x_e}^1 \frac{dx}{x} ~{\hat J}  \\
\times \left[ g_1^\c \Big(2 - \frac{x_e}{x}\Big) 
\left(1 - \frac{Q^2}{2 x_e s}\right)
+ \Big(2g_2^\c + \frac{x_e}{x}g_1^\c\Big) 
\left(1 - \frac{x K^2}{x_e \bar x_e s}\right)\right] \ ,
\label{bm}
\end{multline}
where $\hat J =  \nu_e \bar\nu_e J$, with the Jacobian factor $J$
(which we shall use explicitly in the
section on pseudoscalar meson production) given by
\begin{equation}
J^{-1} = 2 |\epsilon_{\a\b\c\d}p_1^\a p_2^\b q^\c k^\d| =
s k_T q_T \sin\phi \ ,
\label{bn}
\end{equation}
or alternatively,
\begin{equation}
J^{-2} =  K^2 Q^2(s - 2\nu_e)(s- 2\bar\nu_e)  
-4\left(\tfrac{1}{2}s\nu
- \nu_e \bar\nu_e - \tfrac{1}{4}K^2 Q^2\right)^2  \ .
\label{bo}
\end{equation}
Note that the first term in eq.(\ref{bm}) is the same as before, while
the factor
\begin{equation}
\left(1 - \frac{x K^2}{x_e \bar x_e s} \right) = 
\left(1 - \frac{2\nu_e\bar\nu_e}{s\nu}\right) = 
\frac{1}{s\nu} \left(s k_T q_T \cos\phi + \tfrac{1}{2} K^2 Q^2\right)
\label{bp}
\end{equation}
encodes the azimuthal dependence and allows $g_2^\c$ to be determined
from the differential cross-section asymmetry.

\vskip0.3cm
\noindent{\bf (iii)~~Operator product expansion and sum rules:}

In the deep-inelastic limit, the Green function $T_{\m\n\l\r}(q,k)$
can be evaluated using the usual OPE for two electromagnetic currents,
\begin{multline}
i J_\m(q) J_\n(-q) = -i \epsilon_{\m\n\a\s}q^\a \sum_{n=1,~ {\rm odd}}
\left(\frac{2}{Q^2}\right)^n q_{\m_2} \ldots q_{\m_n} \\
\times \left[\sum E_{2,n}(Q^2) R_{2,n}^{\s\m_2 \ldots \m_n}(0) +
\sum E_{3,n}(Q^2) R_{3,n}^{\s\m_2 \ldots \m_n}(0) \right] \ ,
\label{bq}
\end{multline}
where $R_{2,n}$ and $R_{3,n}$ are respectively twist 2 and twist 3 and
we have shown only the odd-parity operators,\footnote{
The sum $\sum$ in eq.\eqref{bq} is over the full set of operators, which
comprises flavour singlet and non-singlet
quark bilinears together with photon and gluon operators (see 
e.g.~refs.\cite{Kodaira:1994ge,Baba:2002mx,Narison:1992fd} 
for a full list). For example,
the singlet quark operators are the symmetric twist 2:
\begin{align*}
R_{2,n}^{\s\m_2 \ldots \m_n} & = i^{n-1} {\cal S} \left[
\bar\psi \c_5 \c^\s D^{\m_2} \ldots D^{\m_n} \psi \right] \\
& \equiv i^{n-1} \frac{1}{n} \bigg[\bar\psi \c_5 \c^\s D^{\m_2} 
 \ldots D^{\m_n} \psi 
+ \sum_{j=2}^n \bar\psi \c_5 \c^{\m_j} D^{\m_2} \ldots
D^\s \ldots D^{\m_n} \psi \bigg] \ ,
\end{align*}
and the antisymmetric twist 3:
\begin{align*}
R_{3,n}^{\s\m_2 \ldots \m_n} & = i^{n-1} {\cal A} \left[
\bar\psi \c_5 \c^\s D^{\m_2} \ldots D^{\m_n} \psi \right] \\
& \equiv i^{n-1} \frac{1}{n} \bigg[(n-1)\bar\psi \c_5 \c^\s D^{\m_2} 
 \ldots D^{\m_n} \psi 
- \sum_{j=2}^n \bar\psi \c_5 \c^{\m_j} D^{\m_2} \ldots
D^\s \ldots D^{\m_n} \psi \bigg] \ ,
\end{align*}
where symmetrisation over the indices $(\m_2,\ldots \m_n)$ is
understood.} which contribute to $g_1^\c$ and $g_2^\c$.
Form factors $\hat R_{2,n}(K^2)$ and $\hat R_{3,n}(K^2)$ are then
defined as: 
\begin{align}
\langle 0| R_{2,n}^{\s\m_2 \ldots \m_n}(0) A_\l(k) A_\r(-k) |0\rangle
& = \frac{i}{K^4} {\cal S} \left[ \hat R_{2,n}(K^2) k^{\m_2} \ldots k^{\m_n}
\epsilon^\d{}_{\b\l\r} k^\b \right] -  {\rm traces} \\ 
\langle 0| R_{3,n}^{\s\m_2 \ldots \m_n}(0) A_\l(k) A_\r(-k) |0\rangle
& = \frac{i}{K^4} {\cal A} \left[ \hat R_{2,n}(K^2) k^{\m_2} \ldots k^{\m_n}
\epsilon^\d{}_{\b\l\r} k^\b \right] -  {\rm traces} \ .
\label{br}
\end{align}

With these definitions, we find \cite{Baba:2002mx}
\begin{multline}
T_{\m\n\l\r}(q,k) = \sum_{n=1, {\rm odd}} \frac{1}{K^4} 
\left(\frac{2}{Q^2}\right)^n \nu^{n-2}   \bigg[
\sum E_{2,n}(Q^2) \hat R_{2,n}(K^2) \left(I_{\m\n\l\r} -
  \frac{n-1}{n} J_{\m\n\l\r}\right)  \\
+ \sum E_{3,n}(Q^2) \hat R_{3,n}(K^2) \frac{n-1}{n} J_{\m\n\l\r} 
\bigg] \ ,
\label{bs}
\end{multline}
and therefore identify the structure functions as:
\begin{align}
g_1^\c(x,Q^2;K^2) & = \frac{2}{\pi} \IM \sum_{n=1, ~{\rm odd}} 
\sum E_{2,n}(Q^2) \hat R_{2,n}(K^2) x^{-n} \\
g_2^\c(x,Q^2;K^2) & = \frac{2}{\pi} \IM \sum_{n=1, ~{\rm odd}}
\frac{n-1}{n} \left[\sum E_{3,n}(Q^2) \hat R_{3,n}(K^2) -
\sum E_{2,n}(Q^2) \hat R_{2,n}(K^2) \right] x^{-n} 
\label{bt}
\end{align}

The moment sum rules follow immediately. For the first structure
function, we have
\begin{equation}
\int_0^1 dx~x^{n-1} g_1^\c(x,Q^2;K^2) = 
\sum E_{2,n}(Q^2) \hat R_{2,n}(K^2) \ .
\label{bu}
\end{equation}
The important first moment sum rule is then
\begin{equation}
\int_0^1 dx~ g_1^\c(x,Q^2;K^2) = 
\sum_{r=3,8,0} E^r_{2,1}(Q^2) \hat R^r_{2,1}(K^2) \ ,
\label{bv}
\end{equation}
where the only operators contributing for $n=1$ are the axial currents 
$J_{\m 5}^r$, with form factors defined from the three-current AVV
Green function:
\begin{equation}
4\pi\a  \langle 0|J_{\m 5}^r(0) J_\l(k) J_\r(-k)|0\rangle = 
i \epsilon_{\l\r\m\a} k^\a \hat R_{2,1}^r(K^2) \ .
\label{bw}
\end{equation}

For the second structure function, we have the moment sum rule
\cite{Baba:2002mx}
\begin{equation}
\int_0^1 dx~x^{n-1}  g_2^\c(x,Q^2;K^2) = \frac{n-1}{n}
\left[\sum E_{3,n}(Q^2) \hat R_{3,n}(K^2)  -
\sum E_{2,n}(Q^2) \hat R_{2,n}(K^2) \right] \ .
\label{bx}
\end{equation}
It follows immediately that the first moment of $g_2^\c$ vanishes,
\begin{equation}
\int_0^1 dx~ g_2^\c(x,Q^2;K^2) = 0 \ .
\label{by}
\end{equation}
This is the Burkhardt-Cottingham sum rule \cite{Burkhardt:1970ti}. 
Following ref.\cite{Baba:2002mx}, we can also isolate the 
contribution of the twist 3
operators. If we define the function $\bar g_2^\c$ as the function
whose moments are given by the twist 3 terms on the r.h.s.~of 
eq.\eqref{bx}, it is straightforward to show that
\begin{equation}
\bar g_2^\c = g_2^\c + g_1^\c - \int_x^1
\frac{dx'}{x'}~g_1^\c(x',Q^2;K^2) \ .
\label{bz}
\end{equation}
The second two terms, which therefore represent (minus) the twist 2
contribution to $g_2^\c$, reproduce the Wandzura-Wilczek 
\cite{Wandzura:1977qf} relation.

For our purposes, we have therefore shown how a measurement of the
azimuthal dependence of the differential cross-section asymmetry at a
polarized $e^+ e^-$ collider such as Super B enables the photon
structure function $g_2^\c$ to be measured as well as $g_1^\c$. The
relation \eqref{bz} then provides a theoretically clean decomposition
allowing the contribution of the twist 3 operators in the OPE to be
isolated and studied in detail.

\newpage

\section{First Moment Sum Rule for $g_1^\gamma(x,Q^2;K^2)$ }

The most interesting direct QCD measurement that could be made at a
high-luminosity polarized $e^+e^-$ collider is the first moment sum
rule for $g_1(x,Q^2;K^2)$.  As shown above, this probes the important 
anomalous 3-current AVV Green function, which encodes a wealth of
information on $U(1)_A$ physics, gluon topology and the realisation 
of chiral symmetry (for a review, see ref.~\cite{Shore:2007yn}).

There are two elements to the sum rule \eqref{bv}. First are the Wilson 
coefficients which are well-known in perturbative QCD and are given to
$O(\a_s)$ by 
\begin{align}
E_{2,1}^r & = c^{(r)} \left(1 - \frac{\a_s(Q^2)}{\pi}\right)\ , \hskip2cm
r = 3,8  \nonumber \\
E_{2,1}^0 & = c^{0)} \left(1 - \frac{\a_s(Q^2)}{\pi}\right) 
\exp\left[\int_0^t dt' \c\left(\a_s(t')\right)\right]\ ,
\label{ca}
\end{align}
where $t = \frac{1}{2}\log\frac{Q^2}{\mu^2}$. For $N_c=3$ and 
with $N_f = 3$ effective dynamical flavours, where $r = 3,8$ labels the
$SU(3)_f$ generators and $r=0$ the singlet, the coefficients $c^{(r)}$
are determined by the quark charges: $c^{(3)} = \frac{1}{3}$,
~$c^{(8)} = \frac{1}{3\sqrt{3}}$ and $c^{(0)} = \frac{2}{9}$.
The flavour singlet current is not conserved because of the $U(1)_A$
anomaly, which gives rise to the non-vanishing anomalous dimension 
$\c = - \c_0 \frac{\a_s}{4\pi} - \c_1 \frac{\a_s^2}{(4\pi)^2} -
\ldots$ with $\c_0= 0$ and $\c_1 = 6N_f (N_c^2-1)/N_c = 48$. 
It is important to note that 
this expansion only starts at $O(\a_s^2)$. We also need the beta
function: $\b = - \b_0 \frac{\a_s^2}{4\pi} - \b_1 \frac{\a_s^3}{(4\pi)^2} -
\ldots$ with $\b_0 = \frac{2}{3}(11N_c -2N_f) = 18$.

The second element is the AVV Green function itself. In general, this
is given in terms of a set of six form factors $A_1, \ldots, A_6$ by
\begin{align}
-i \langle 0|J_{\mu 5}^r(p) J_\l(k_1) J_\r(k_2)|0\rangle & = 
A_1^r \epsilon_{\m\l\r\a}k_1^\a + A_2^r \epsilon_{\m\l\r\a}k_2^\a 
\nonumber \\
& + A_3^r \epsilon_{\m\l\a\b}k_1^\a k_2^\b k_{2\r} 
+ A_4^r \epsilon_{\m\r\a\b}k_1^\a k_2^\b k_{1\l} \nonumber \\
& + A_5^r \epsilon_{\m\l\a\b}k_1^\a k_2^\b k_{1\r} 
+ A_6^r \epsilon_{\m\r\a\b}k_1^\a k_2^\b k_{2\l}  \ ,
\label{cb}
\end{align}
where the form factors are functions of the invariant momenta:
$A_1^r = A_1^r(p^2,k_1^2,k_2^2)$, etc. For simplicity, we abbreviate
$A_i^r(0,k^2,k^2) = A_i^r(K^2)$ below. The form factor 
$\hat R_{2,1}^r(K^2)$ in the first moment sum rule \eqref{bv}
is therefore simply
\begin{equation}
\hat R_{2,1}^r(K^2) = 4\pi\a \left(A_1^r(K^2) - A_2^r(K^2)\right)\ .
\label{cc}
\end{equation}

The $K^2=0$ limit of $\hat R_{2,1}^r$ is determined by electromagnetic
current conservation \cite{Bass:1991sg,Narison:1992fd,Shore:1992pm,
Bass:1998bw}. 
The Ward identity $\partial^\l J_\l= 0$ applied
to the AVV function implies
\begin{equation}
ik_1^\l\langle0|J_{\m 5}^r(p) J_\l(k_1) J_\r(k_2)|0\rangle = 0 \ ,
\label{cd} 
\end{equation}
and similarly for $k_1\rightarrow k_2$. Substituting the form factor
decomposition, we find
\begin{align}
A_1^r &  = A_3^r k_2^2 + A_5^r \frac{1}{2}(p^2 - k_1^2 -k_2^2)
\nonumber\\
A_2^r & = A_4^r k_1^2 + A_6^r \frac{1}{2}(p^2 - k_1^2 -k_2^2) \ ,
\label{ce}
\end{align}
so in the limit $p\rta 0$, $k_1^2 = k_2^2 = -K^2$, the r.h.s.~vanishes
at $K^2 \rta 0$ provided none of the form factors is singular, as is
the case in the absence of exactly massless Goldstone bosons
coupling to $J_{\m5}^r$. It follows immediately that 
$\hat R_{2,1}^r(0) = 0$, and so the first moment of $g_1^\c$ for real
photons vanishes:
\begin{equation}
\int _0^1 dx~g_1^\c(x,Q^2;K^2=0) = 0 \ .
\label{cf}
\end{equation}

The asymptotic limit for large $K^2$ (while still retaining the DIS
condition that $Q^2 \gg K^2$) can be deduced using the renormalization
group together with the anomalous chiral Ward identity for 
$J_{\m 5}^r$. This is:
\begin{equation}
\partial^\m J_{\m 5}^r = d_{rst} m_s \phi_5^t + 6 Q \d^{r0} 
+ a^{(r)} \frac{\a}{8\pi} \tilde F^{\m\n}F_{\m\n} \ ,
\label{cg}
\end{equation}
where $\phi_5^r = \bar\psi T^r \c_5 \psi$ and $Q = 
\frac{\a_s}{8\pi} {\rm tr} \tilde G^{\m\n} G_{\m\n}$ is the
topological charge density.  $G_{\m\n}$ and $F_{\m\n}$ are the gluon
and electromagnetic field strengths, and the quark masses are written
in $SU(3)_f$ notation as ${\rm diag}(m_u,m_d,m_s) = 
\sum_{r=3,8,0} m_r T^r$ with $d_{rst}$ the usual $d$-symbols.
The gluonic anomaly term involving $Q$ arises only for the $U(1)_A$
flavour singlet current $J_{\m 5}^0$  while the final term is the
usual electromagnetic axial anomaly, with coefficients
$a^{(3)} = 1$, $a^{(8)}= \frac{1}{\sqrt3}$ and $a^{(0)} = 4$ 
determined by the quark charges. The AVV Green function therefore
satisfies 
\begin{multline}
i p^\m \langle0|J_{\m 5}^r(p) J_\l(k_1) J_\r(k_2)|0\rangle = 
d_{rst}m_s \langle0|\phi_5^r(p) J_\l(k_1) J_\r(k_2)|0\rangle \\
+ 6 \d^{r0} \langle0|Q(p) J_\l(k_1) J_\r(k_2)|0\rangle
+ a^{(r)} \frac{1}{8\pi^2} \e_{\l\r\a\b}k_1^\a k_2^\b \ ,
\label{ch}
\end{multline}
which in the limit $p\rta0$ implies
\begin{equation}
A_1^r(K^2) - A_2^r(K^2) = D^r(K^2) + B^r(K^2) + \frac{1}{8\pi^2}
a^{(r)} \ ,
\label{ci}
\end{equation}
where $D^r$ and $B^r$ are form factors defined in the obvious way 
from the Green functions involving $\phi_5^r$ and $Q$.
To determine the large $K^2$ behaviour, note that the form factors 
$A_i^r(K^2)$ for the flavour non-singlet currents satisfy a homogeneous RG
equation with the standard solution in terms of running couplings and
masses, while for the flavour singlet there is an additional anomalous
dimension contribution:
\begin{align}
A_i^r(K^2;\a_s(\m);m) & = A_i^r\big(\m^2;\a_s(t);e^{-t}m(t)\big)
\hskip 3cm r=3,8 \nonumber \\
A_i^0(K^2;\a_s(\m);m) & = A_i^0\big(\m^2;\a_s(t);e^{-t}m(t)\big)
\exp\left[-\int_0^t dt'~\c(\a_s(t'))\right] \ ,
\label{cj}
\end{align}
where here $t= \frac{1}{2}\log\frac{K^2}{\m^2}$. Similar results
hold for $D^r(K^2)$ and $B^r(K^2)$.  Clearly, therefore, in the limit
$K^2\rta \infty$ the mass term goes to zero and $D^r(K^2)$ does not
contribute. Moreover, for small $\a_s(t)$, the contribution $B^r(K^2)$ 
of the topological charge term, which is $O(\a_s^2)$, can be
neglected at the NLO order we are working to here. (See \cite{Sasaki:2006bt}
for a discussion of the sum rule at $O(\a_s^2\a)$.)
We therefore deduce
\begin{align}
\hat R_{2,1}^r(K^2\rta\infty) & = \frac{1}{2} a^{(r)} \frac{\a}{\pi}
\hskip3cm r=3,8   \nonumber \\
\hat R_{2,1}^0(K^2\rta\infty) & = \frac{1}{2} a^{(0)} \frac{\a}{\pi}
\exp\left[-\int_0^t dt'~\c(\a_s(t'))\right] \ .
\label{ck}
\end{align}
The asymptotic form of the sum rule then follows by combining 
eq.\eqref{ca} for the Wilson coefficients with eq.\eqref{ck}
for the form factors. We find \cite{Narison:1992fd}:
\begin{multline}
\int_0^1 dx~g_1^\c(x,Q^2;K^2\rta\infty) =
\sum_{r=3,8,0} E_{2,1}^r(Q^2) \hat R_{2,1}^r(K^2) \\
= \frac{1}{2} \frac{\a}{\pi} \left(1 - \frac{\a_s(Q^2)}{\pi}\right)
\left[c^{(3)} a^{(3)} + c^{(8)} a^{(8)} + c^{(0)} a^{(0)} 
\exp\bigg[\int_{t(K)}^{t(Q)} dt'~\c(\a_s(t'))\bigg] \right] \ .
\label{cl}
\end{multline}
Finally, using $\frac{\a_s(t)}{4\pi} \simeq \frac{1}{\b_0 t}$ and
substituting for the $c^{(r)}$ and $a^{(r)}$ coefficients, we 
obtain the result for $N_c= N_f = 3$ QCD at $O(\a_s \a)$
quoted in the introduction \cite{Narison:1992fd}:
\begin{equation}
\int_0^1 dx~g_1^\c(x,Q^2;K^2\rta\infty) =
\frac{2}{3} \frac{\a}{\pi} \left(1 - \frac{\a_s(Q^2)}{\pi}\right)
\left[1 + \frac{4}{9} \left(\frac{\a_s(Q^2)}{\pi} -
\frac{\a_s(K^2)}{\pi}\right)\right] \ .
\label{cm}
\end{equation}
Note that the overall coefficient is $N_c \sum_f \hat e_f^2$, 
i.e.~proportional to the fourth power of the quark charges 
$\hat e_f$ as given by the lowest-order box diagram contributing to
$g_1^\c$.

For intermediate values of $K^2$, we may rewrite the sum rule in the
convenient form
\begin{multline}
\int_0^1 dx~g_1^\c(x,Q^2;K^2) = \frac{1}{18} \frac{\a}{\pi}
\left(1 - \frac{\a_s(Q^2)}{\pi}\right) \\
\times \left[3F^3(K^2) + F^8(K^2) + 8F^0(K^2;\m^2=K^2)
\exp\bigg[\int_{t(K)}^{t(Q)} dt'~\c(\a_s(t'))\bigg]\right] \ ,
\label{cn}
\end{multline}
where we have introduced normalised form factors $F^r(K^2)$ 
defined by
\begin{equation}
A_1^r(K^2) - A_2^r(K^2) = \frac{1}{8\pi^2} a^{(r)} F^r(K^2) \ .
\label{co}
\end{equation}
Note that in anomalous flavour singlet sector, with the choice
of anomalous dimension factor in eq.\eqref{co}, the 
renormalization scale in $F^0(K^2;\m^2)$ is specified as $\m^2 = K^2$.

The form factors $F^r(K^2)$ therefore interpolate between
0 for $K^2=0$ and 1 for asymptotically large $K^2$. 
The full momentum dependence of the sum rule for $g_1(x,Q^2;K^2)$
is governed by these form factors, which are in turn determined by the
AVV Green finction.  A non-perturbative, first-principles calculation of this
3-current Green function would therefore give a complete prediction
for the first moment sum rule for arbitrary photon virtuality $K^2$.

In practice, this is still beyond current techniques and represents a
challenge to lattice gauge theory, QCD spectral sum rules, AdS/QCD 
and other non-perturbative approaches to QCD. 
Indeed, this emphasises the importance of a direct experimental 
measurementof the sum rule and the form factors $F^r(K^2)$.
As an interim measure we can adopt a phenomenological approach,
modelling the form factor by a simple interpolating formula such as
$F^r(K^2) \simeq K^2/(K^2 + M^2)$ for some characteristic crossover
scale $M^2$. For heavy quarks, this would be the quark
mass itself. However, for the light quarks, due to chiral symmetry
breaking, we expect $M^2$ to be a typical hadronic scale, viz. 
$M^2 \sim m_\r^2$ for $F^3(K^2)$.  This can be
supported by a simple OPE argument
\cite{Narison:1992fd,Shore:2004cb}. 
If we evaluate the
3-current Green function by inserting an intermediate pseudoscalar
meson, then write the OPE for the remaining electromagnetic currents
for large $K^2$, we have
\begin{equation}
\langle \pi|J_\l(k) J_\r(-k)|0\rangle = -\frac{2}{K^2} \e_{\l\r\a}{}^\m k^\a
E_{2,1}^3(K^2) \langle \pi|J_{\m 5}^3(0)|0\rangle + \ldots \ ,
\label{cp}
\end{equation}
which implies
\begin{equation}
F^3(K^2\rta\infty) = 1 - \frac{(4\pi)^2}{3}f_\pi^2 \frac{1}{K^2} +
\ldots 
\label{cq}
\end{equation}
The crossover scale is then identified from this expansion as
$M^2 \simeq \frac{(4\pi)^2}{3}f_\pi^2 \sim m_\r^2$, as would
be expected in general terms from vector meson dominance.\footnote{
The VMD analysis has been carried through in detail in ref.\cite{Ueda:2006cp}
to obtain a numerical estimate of $g_1^\c(x,Q^2;K^2)$ in the
non-perturbative region. Essentially, VMD involves evaluating 
the AVV correlation function by replacing the electromagnetic
currents with the corresponding vector mesons
$\r, \omega$ and $\phi$. In our notation, ref.\cite{Ueda:2006cp}
quotes the following formula for the off-shell form factors:
\begin{equation*}
F^r(K^2) = 1 - \sum_{V=\r,\omega,\phi} c_V
\left(\frac{m_V^2}{K^2 + m_V^2}\right)^2 
= \sum_{V=\r,\omega,\phi} c_V
\frac{K^2(K^2 + 2m_V^2)}{(K^2 + m_V^2)^2}  
\end{equation*}
where the couplings $c_V\sim1/f_V^2$ are constrained by 
$\sum_V c_V= 1$.
Note however that this gives a different high $K^2$ behaviour
$F^r(K^2) \sim 1 + O(m_V^4/K^4)$ from the OPE estimate
above. See also the discussion around eq.~(190) of 
chapter 5 in ref.~\cite{Jegerlehner:2009ry} and
ref.\cite{Melnikov:2003xd}, where this is discussed in the
context of the hadronic light-by-light contributions to the 
muon $g-2$.}

\newpage

\section{Two-photon physics and pseudoscalar mesons }

In this section, we consider other aspects of polarized two-photon
physics accessible at Super B, with a special focus on the pseudoscalar
mesons $P = \pi,\eta,\eta'$  and their radiative transition functions
$g_{P\c\c}(k_1^2,k_2^2)$. As we shall see, these may be measured for
complementary values of the photon invariant momenta $k_1^2$ and 
$k_2^2$ either from the form factors arising in the $g_1^\c$ moment
sum rule or by direct two-photon production of the pseudoscalar
mesons.

\vskip0.3cm
\noindent{\bf (i)~~Pseudoscalar mesons and the $g_1^\c(x,Q^2;K^2)$ 
sum rule:}

In the first approach, we use familiar PCAC ideas to rewrite the form
factors characterising the 3-current AVV function in terms of the
pseudoscalar meson transition functions by assuming the dominant
contribution comes from the pseudo-Goldstone boson intermediate
states.\footnote{Notice, for example, that this actually gives the transition
  function $g_{\pi\c\c}$ for zero pion momentum. The extrapolation to
  the physical transition function for on-shell pions is assumed to be
  smooth, in the spirit of conventional applications of PCAC to light,
  though not massless, pseudo-Goldstone bosons. The same smoothness
  assumption is also made for the heavier $\eta$ and $\eta'$.}  This is not
straightforward, as careful account has to be taken of $SU(3)_f$
mixing and, especially, the interesting and subtle use of PCAC in the
anomalous $U(1)_A$ channel. This has been described in detail in our
earlier work \cite{Narison:1992fd,Shore:1992pm,Shore:1991np,
Shore:1999tw,Shore:2006mm}, 
and specifically in ref.\cite{Shore:2004cb} where the following
results were presented:
\begin{align}
& f_{\pi}g_{\pi\c^*\c^*}(K^2,K^2)  = \frac{\a}{\pi} \left(1 - F^3(K^2)\right)
\nonumber\\
&f_{8\eta}g_{\eta\c^*\c^*}(K^2,K^2) + f_{8\eta'}g_{\eta'\c^*\c^*}(K^2,K^2)
 = \frac{1}{\sqrt3} \frac{\a}{\pi} \left(1 - F^8(K^2)\right)
\nonumber\\
&f_{0\eta}g_{\eta\c^*\c^*}(K^2,K^2)  + f_{0\eta'}g_{\eta'\c^*\c^*}(K^2,K^2) 
+ 6A g_{G\c^*\c^*}(K^2,K^2;\m^2)  =
4 \frac{\a}{\pi} \left(1 - F^0(K^2;\m^2)\right)
\label{da}
\end{align} 

The flavour singlet relation \eqref{da} is particularly interesting
theoretically, since it involves the non-perturbative constant $A$
which determines the gluon topological susceptibility in 
QCD \cite{Crewther:1979pi,Leutwyler:1992yt}:
\begin{equation}
\chi(0) \equiv \langle~Q~ Q~\rangle = -A \left[1 - A \sum_q
\frac{1}{m_q\langle\bar q q\rangle}\right]^{-1} \ .
\label{db}
\end{equation}
The corresponding transition function $g_{G\c^*\c^*}$ determines the
coupling of two photons to a glueball-like operators $G$ which is
orthogonal to the physical $\eta'$. However, $G$ does not necessarily
correspond to a physical particle state so is not directly measurable.
In fact, theoretical arguments based on the $1/N_c$ expansion show
that the $6Ag_{G\c^*\c^*}$ term in \eqref{db} is sub-dominant, while an
explicit fit of the transition functions and $SU(3)_f$-mixed decay 
constants (see ref.\cite{Shore:2006mm} for full details and experimental
values) shows that its relative contribution is likely to be only
around 20\%. 

Setting this important subtlety to one side, we may therefore
determine the off-shell pseudoscalar meson transition functions
$g_{\pi\c^*\c^*}$, $g_{\eta,\c^*\c^*}$ and $g_{\eta'\c^*\c^*}$ in the kinematical
region where the photon invariant momenta are equal and cover the full
range of $K^2$ accessible to the experimental measurement of the
structure function $g_1^\c(x,Q^2;K^2)$.

\vskip0.3cm
\noindent{\bf (ii)~~Exclusive two-photon production and meson
  transition functions:}

The second approach involves the direct measurement of the transition
functions for the pseudoscalar mesons $P = \pi,\eta,\eta'$ through the
two-photon production reaction $e^+ e^- \rta e^+ e^- P$ shown in
Fig.~2 \cite{Brodsky:1970vk,Brodsky:1971ud,Brodsky:1971vm}.
(See also ref.~\cite{Kolanoski:1987sn} for an early review of 
two-photon physics at $e^+ e^-$ colliders.)
This gives the transition functions $g_{P\c^*\c^*}(Q^2,K^2)$ where
$Q^2$ and $K^2$ are measured directly provided the scattering angles of
both electrons are tagged. This may cover the whole range from $Q^2$ 
values typical of DIS to soft, nearly-real $K^2$ as well as all
intermediate values.

\begin{figure}[ht] 
\centerline{\includegraphics[width=2.3in]{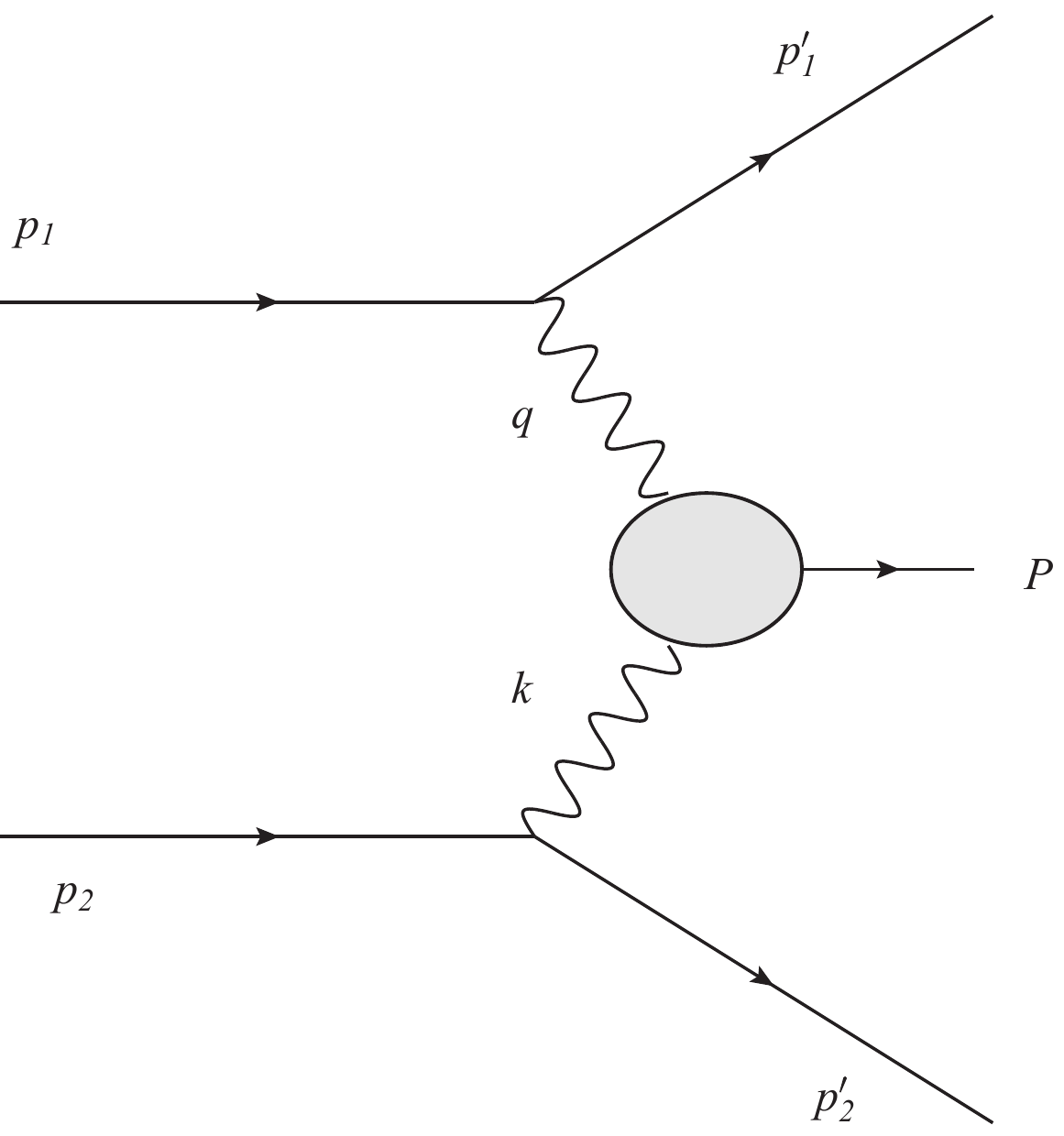}}
\caption{\footnotesize Kinematics for the two-photon pseudoscalar
  meson production reaction $e^+ e^- \rta e^+ e^- P$.}
\label{Fig2}
\end{figure}

The cross-section for $e^+ e^- \rta e^+ e^- P$ is given by 
\begin{multline}
\s = \frac{1}{2}\frac{\a^2}{(2\pi)^4}\frac{1}{s} \int d^4q~ 
\d\left(p_1.q + \tfrac{1}{2}Q^2\right) \int d^4k~
\d\left(p_2.k + \tfrac{1}{2}K^2\right) \frac{1}{Q^4K^4} \\
\times L^{\m\n}(p_1,h_1;q) M^{\dagger}_{\m\l}(q,k) M_{\n\r}(q,k)
L^{\l\r}(p_2,h_2;k) ~2\pi \d(2\n - Q^2 - K^2 -m_P^2) \ ,
\label{dc}
\end{multline}
where 
\begin{equation}
M_{\m\l}(q,k) = -i \e_{\m\l\a\b}q^\a k^\b g_{P\c^*\c^*}(Q^2,K^2)
\label{dd}
\end{equation}
is defined in terms of the off-shell transition functions.\footnote{
The on-shell transition functions are defined from the decays
$P\rta\c\c$ by
\begin{equation*}
 {\cal M}_{(\l_1)( \l_2)}(q,k)=  \langle\c(q) \c(k)|P\rangle =
-i \e_{\m\l\a\b}q^\a k^\b \e_{(\l_1)}^\m(q) \e_{(\l_2)}^\l(k) g_{P\c\c}\ ,
\end{equation*}
where $\e_{(\l_1)}^\m(q)$, $\e_{(\l_2)}^\l(k)$ are the photon
polarization vectors.  With this definition, the decay rate is
\begin{equation*}
\Gamma(P\rta\c\c) = \frac{m_P^3}{64\pi} |g_{P\c\c}|^2 \ .
\end{equation*}
}
This is related to the inclusive cross-section \eqref{bh}
by the optical theorem, which for the specific meson final state $P$
implies the substitution
\begin{equation}
4\pi\a K^4 \IM T_{\m\n\l\r} \rta \frac{1}{2} M^{\dagger}_{\m\l}(q,k)  
M_{\n\r}(q,k)~
2\pi \d\left((q+k)^2 - m_P^2\right) \ .
\label{de}
\end{equation}
Notice that in deriving \eqref{bh} no use has been made of the
conventional `equivalent photon' formalism 
(see, {\it e.g.}~refs.~\cite{Berger:1986ii,Brodsky:1971ud}
and no assumption 
has been made that the target photon is quasi-real. This is essential
if we are to measure transition functions covering the whole range of
values of $K^2$.

We can evaluate the integrand in this expression for the cross-section
in the same way as in section 2. Defining the symmetric part of the
leptonic tensor of \eqref{be} as $L^S_{\m\n}(p_1,q)$, we find
\begin{multline}
L_S^{\m\n}(p_1,q) M^{\dagger}_{\m\l}(q,k)  M_{\n\r}(q,k)
L_S^{\l\r}(p_2,k)  \\
= \Big[4J^{-2} + 4Q^2 K^2\Big(\left(\n \n_e - \tfrac{1}{4}Q^2 K^2\right) - 
\n_e^2 + \left(\n \bar\n_e - \tfrac{1}{4}Q^2 K^2\right) - \bar\n_e^2 
- \tfrac{1}{2}\left(\n^2 - Q^2 K^2\right)\Big) \Big] \\
\times ~|g_{P\c^*\c^*}|^2
\label{df}
\end{multline}
where $J$ is the Jacobian factor in \eqref{bo}. Substituting for $J$,
and after reorganising terms on the r.h.s, we find the comparatively 
simple expression:
\begin{equation}
\left[4 \left(s\n - 2\n_e \bar\n_e\right)^2
- 4Q^2 K^2 \left(\left(s + \tfrac{1}{2}\n - \n_e - \bar\n_e\right)^2 
+ \tfrac{1}{4}(\n^2 - Q^2 K^2) \right) \right]|g_{P\c^*\c^*}|^2 \ .
\label{dg}
\end{equation}
We recognise the first term here as that previously found in
\eqref{bp}, which encodes the dependence on the azimuthal 
scattering angle $\phi$. To determine the polarization asymmetry of
the cross-section, we also require the contribution of the antisymmetric
part of the leptonic tensor:
\begin{multline}
L_A^{\m\n}(p_1,q) M^{\dagger}_{\m\l}(q,k)  M_{\n\r}(q,k)
L_A^{\l\r}(p_2,k) \\
= -4 Q^2 K^2 (\n - 2\n_e) (\n - 2\bar\n_e) ~|g_{P\c^*\c^*}|^2 \ .
\label{dh}
\end{multline}
 
Putting all this together, we find the following 
formula\footnote{This may be compared with the corresponding
  cross-section formula in ref.\cite{Brodsky:1971ud}, where exclusive meson 
  production in unpolarized $e^+e^-$ scattering was first analysed. In
  particular, the two terms in \eqref{dg} can be recognised as $B_2$,
  $B_1$ in eq.(4.6) of ref.\cite{Brodsky:1971ud}.  For polarized scattering, the
  term in \eqref{dh} may be identified as the antisymmetric spin 0
  contribution $\tau_{TT}^a$ to the $\c^* \c^*\rta \c^*\c^*$
  cross-section in ref.\cite{Pascalutsa:2012pr} 
  (see next sub-section), which quotes
  a formula for the $e^ +e^-\rta e^+ e^- X$ cross-section in terms of
  the eight independent helicity amplitudes for light-by-light scattering.}
relating the differential cross-section for the exclusive production 
reaction $e^+ e^- \rta e^+ e^- P$ and the off-shell pseudoscalar 
meson transition functions $g_{P\c*\c*}(Q^2,K^2)$:
\begin{align}
\frac{d^2\s(h_1,h_2)}{dQ^2 dK^2} & = 
\frac{\a^2}{(2\pi)^2}\frac{1}{s^2Q^4 K^4} \int_0^\infty d\n_e 
\int_0^\infty d\bar\n_e \int_0^\infty d\n
~\d\left(\n -\tfrac{1}{2}( Q^2 + K^2 + m_P^2)\right)~J \nonumber\\
&\times \bigg[(s\n - 2\n_e \bar\n_e)^2 - Q^2 K^2\Big(
(s + \tfrac{1}{2}\n - \n_e - \bar\n_e)^2 + 
\tfrac{1}{4}(\n^2 - Q^2 K^2)^2 \nonumber \\
&\hskip3cm~~  - h_1 h_2 (\n - 2\n_e)(\n - 2\bar\n_e)
\Big) \bigg] ~
|g_{P\c^*\c^*}(Q^2,K^2)|^2 \ .
\label{di}
\end{align}
This shows clearly how the off-shell transition functions
$g_{P\c^*\c^*}(Q^2,K^2)$, for the full range of $Q^2$ and $K^2$, may
be extracted from the exclusive differential cross-section.
Notice that in this case, where the produced hadron is a pseudoscalar,
there is only a single transition function (see eq.\eqref{dd}) and so
no extra information is obtained from the polarization asymmetry of
the cross-section. The transition functions can therefore be obtained
from an $e^+e^-$ collider running even with unpolarized beams,
as was the case with BABAR. This is not true of higher-spin mesons, 
where knowing the polarized cross-sections analogous to \eqref{di} 
will yield valuable new information. This is especially relevant 
to $\c^*\c^*\rta\c^*\c^*$ scattering, as discussed below.

Determining the $\pi,\eta,\eta'$ transition functions in this way will
complement other low-energy experimental studies of $\eta$ and
$\eta'$ physics, and add to our understanding of other processes 
such as $\eta(\eta') \rta V\c$, where $V = \r,\w,\phi$ and
vector meson dominance can be tested, 
or $\eta'(\eta) \rta \pi^+ \pi^- \c$. For our purposes,
it will also add to the usefulness of the set of identities \eqref{da}.
An independent, direct measurement of the off-shell transition
functions $g_{P\c^*\c^*}(Q^2,K^2)$ will allow the form factors $F^r(K^2)$
to be determined and used as input into the $g_1^\c(x,Q^2;K^2)$ sum
rule. It will also clarify the role of the anomalous gluonic term in
\eqref{da} and provide indirect experimental information on the gluon
topological susceptibility in QCD.

Here, we are primarily interested in $\c^* \c^*$ reactions where the
target photon is off-shell and both electrons are tagged (see also
\cite{Pire:2006ik}). This complements the extensive programme of 
$\c^* \c$, single-tagged, scattering off quasi-real photons which can
also be carried out at Super-B \cite{PSW}. In addition to 
$\c^* \c \rta P$ at high $Q^2$, which is interpreted \cite{Brodsky:1981rp}
in terms of meson distribution amplitudes, there is considerable
interest in $\c^* \c \rta \pi\pi, \rho\rho$, etc.
\cite{Brodsky:1971ud, Diehl:2000uv, Achard:2003qa}
which can be interpreted in terms of generalised distribution amplitudes
(GDAs), related to the GPDs used to analyse deeply-virtual Compton 
scattering and the angular momentum decomposition of the nucleon.
Reactions producing hybrid mesons \cite{Anikin:2006du}
are also of interest.

\vskip0.3cm
\noindent{\bf (iii)~~Light-by-light scattering:}

Light-by-light scattering is a fundamental quantum process,
interesting both in its own right and because it arises theoretically
in the calculation of the anomalous magnetic moment $g-2$ of the
muon. Indeed, the hadronic contribution to the light-by-light
contribution is currently the major theoretical uncertainty, which in
turn constrains the interpretation of any anomalies in the muon
$g-2$ as a signal of new physics beyond the standard model
\cite{Jegerlehner:2009ry}.

The imaginary part of the light-by-light scattering amplitude is related
via the optical theorem to the related process of $\c\c$ fusion. 
As discussed above,
cross-sections for $\c^*\c^*\rta X~{\rm hadrons}$ can be readily
measured in $e^+ e^-$ colliders and a high-luminosity, polarized
machine such as Super B is ideally suited for this purpose. 

Labelling the photon helicities by $\l_1$,$\l_2$, the imaginary part
of the virtual  light-by-light forward scattering amplitude
is written (compare eq.\eqref{de}) as 
\begin{multline}
\IM M_{(\l_1)(\l_2),(\l'_1)(\l'_2)}(q,k) = \frac{1}{2} \sum_X \int d\Gamma_X
  ~(2\pi)^4 \d(q+k-p_X)~\\
 {\cal M}^\dagger_{(\l_1)(\l_2)}(q,k;p_X)
{\cal M}_{(\l'_1)(\l'_2)}(q,k;p_X) \ ,
\label{dj}
\end{multline}
where ${\cal M}_{(\l_1)(\l_2)}(q,k;p_X)$  is the
amplitude for $\c^*(q;\l_1) + \c^*(k;\l_2) \rta X(p_X)$,
with $X$ denoting a hadronic state and $d\Gamma_X$ the
corresponding phase space measure. There are a total of eight
independent helicity amplitudes in 
$\IM M_{(\l_1)(\l_2),(\l'_1)(\l'_2)}$, which can be related to the
cross-sections for $\c^*\c^*\rta X$ with different photon 
helicities and spins of the hadronic state $X$. These are listed in
full in ref.\cite{Pascalutsa:2012pr}, together with the corresponding dispersion
relations. In particular, to make contact with the results above,
if the hadron is a pseudoscalar meson $P$ then the only non-vanishing
scattering amplitudes ${\cal M}_{(\l_1)(\l_2)}$ are those with the
photons transversely polarized in the same sense, and we can show
from the definition in footnote 8 that
\begin{equation}
{\cal M}_{++}(q,k) = - {\cal M}_{--}(q,k) = - 4\pi\a \left(\n^2 -
  Q^2K^2\right)^{\tfrac{1}{2}} ~ g_{P\c^*\c^*}(Q^2,K^2) \ . 
\label{dk}
\end{equation}
The other light-by-light helicity amplitudes 
$M_{(\l_1)(\l_2)(\l'_1)(\l'_2)}(q,k)$ are similarly determined from
measurements of the $\c^*\c^*$ amplitudes 
${\cal M}_{(\l_1)(\l_2)}(q,k;p_X)$, which are in turn measured
from the exclusive $e^+e^- \rta e^+e^-X$ cross-sections for different
hadronic states and electron polarizations.

So once more we see from a different perspective the importance of
measuring the off-shell meson transition functions. 
In particular, we want to highlight the close connection between the
form factors describing low-energy pseudoscalar meson decays, the
light-by-light scattering amplitudes relevant to the muon $g-2$, 
and the first moment sum rule for the photon structure functions
$g_1^\c(x,Q^2;K^2)$. This whole rich spectrum of complementary QCD 
phenomena would be experimentally accessible given a dedicated 
programme of two-photon physics at Super B.

\section{QCD at Super-B}

In this final section, we use the design parameters of Super-B
to investigate the feasibility of the programme of two-photon QCD
physics described above, with particular focus on the possibility of
verifying the first moment sum rule for $g_1^\c(x,Q^2;K^2)$.
Here, the key issues concern the luminosity, beam polarization 
and the possibility of tagging the target electron.

The first issue is whether the luminosity is sufficiently high to
allow $g_1^\c(x,Q^2;K^2)$ to be measured from the polarization
asymmetry of the differential cross section, according to
eq.\eqref{bl}. The total, polarization averaged, cross-section
$\s$ is expressed in terms of the photon structure functions 
$F_2^\c(x,Q^2;K^2)$ and $F_L^\c(x,Q^2;K^2)$ by
\begin{multline}
\s = {\a^3} \int_0^\infty\frac{dQ^2}{Q^4}
\int_0^1\frac{dx_e}{x_e} \int_0^\infty\frac{dK^2}{K^2}
\int_{x_e}^1\frac{dx}{x} ~\frac{x_e}{x}
P_{\c e}\left(\frac{x_e}{x}\right) \\
\times \left[F_2^\c\left(1 - \frac{Q^2}{x_e s}
+ \frac{1}{2} \frac{Q^4}{x_e^2 s^2}\right)
- \frac{1}{2} F_L^\c \frac{Q^4}{x_e^2 s^2} \right]
\label{ea} 
\end{multline}
where $P_{\c e}(z) = (1 + (1-z)^2)/z$.  To find an initial estimate, 
we again work to leading order in $Q^2/x_e s$, so retain only the 
$F_2^\c$ contribution.  The polarization asymmetry is given by 
\eqref{bj},
\begin{equation}
\D\s = {\a^3\over s} \int_0^\infty{dQ^2\over Q^2} \int_0^1{dx_e\over
  x_e}  \int_0^\infty{dK^2\over K^2} \int_{x_e}^1 {dx\over x} ~
\D P_{\c e}\left({x_e\over x}\right) ~g_1^\c(x,Q^2;K^2) ~
\Big(1 - {Q^2\over 2x_e s}\Big) \ .
\label{eb}
\end{equation}

To estimate these cross-sections, we use the dominant contribution
to the moments, viz.
\begin{align}
\int_0^1dx~ x^{n-1} F_2^\c(x,Q^2;K^2) & \simeq {\a\over4\pi} a_{n+1}
\log\frac{Q^2}{\L^2}\ , \hskip1cm  n\ge 1,~{\rm odd} \nonumber\\
\int_0^1dx~ x^{n-1} g_1^\c(x,Q^2;K^2) & \simeq {\a\over4\pi} b_{n+1}
\log\frac{Q^2}{\L^2}\ , \hskip1cm  n\ge 3,~{\rm odd} \ ,
\label{ec}
\end{align}
where $a_{n+1}, b_{n+1}$ are known 
\cite{Bardeen:1978hg, Manohar:1988pp}.
Taking the inverse Mellin transform, we deduce
\begin{align}
F_2^\c & \simeq \frac{\a}{4\pi} a(x) \log\frac{Q^2}{\L^2} \nonumber\\
g_1^\c & \simeq \frac{\a}{4\pi} b(x) \log\frac{Q^2}{\L^2} \ ,
\label{ed}
\end{align}
where numerically $a(x)$ and $b(x)$ are approximately constant, with
$1.2 < a(x) <1.6$ for $0.3 < x < 0.9$. 
For the estimates below, we take $a(x) \simeq \bar a \simeq 1.5$ with
the same value for $b(x) \simeq \bar b$. Placing upper and lower
limits on the integrations over $Q^2$, $K^2$, $x_e$ and $x$, we
therefore find \cite{Narison:1992fd}:
\begin{align}
\s & \simeq \frac{\a^4}{2\pi} \bar a \frac{1}{Q^2_{min}} 
\log\frac{Q^2_{min}}{\L^2} \log\frac{K^2_{\max}}{K^2_{min}}
\log\frac{x_e^{max}}{x_e^{min}} \log\frac{x_{max}}{\langle x_e\rangle}
\nonumber\\
\Delta\s & \simeq \frac{\a^4}{2\pi} \bar b \frac{1}{s}
\log\frac{Q^2_{max}}{Q^2_{min}} \log\frac{\langle Q^2\rangle}{\L^2}
\log\frac{K^2_{\max}}{K^2_{min}}
\log\frac{x_e^{max}}{x_e^{min}} \log\frac{x_{max}}{\langle x_e\rangle}\ ,
\label{ee}
\end{align}
where $\langle Q^2\rangle$ is the geometric mean of $Q^2_{max}$ and
$Q^2_{min}$, and similarly for $x_e$.  The ratio $\Delta\s/\s$ is 
therefore 
\begin{equation}
\frac{\Delta\s}{\s} \simeq \frac{1}{2} \frac{Q^2_{min}}{s} 
\log\frac{Q^2_{max}}{Q^2_{min}} \left[1 + \log\frac{Q^2_{max}}{\L^2}
\left(\log\frac{Q^2_{min}}{\L^2}\right)^{-1} \right] \ .
\label{ef}
\end{equation}

The experimental cuts are chosen as follows.
We take $K^2_{min}\simeq 0.1~ {\rm GeV}^2$ and
$K^2_{max} \simeq 2~{\rm GeV}^2$, above the non-perturbatively
interesting region $K^2 \sim m_{\r}^2$ where the first moment 
of $g_1^\c(x,Q^2;K^2)$ rises from 0 to its asymptotic value.
For the Bjorken variables, we choose $\n_e^{max}=\n_{max} \simeq s/2$
from \eqref{ba}, and $\n_e^{min}=\n_{min} \simeq Q^2_{min}/2$
to ensure $x_e$ and $x$ greater than zero, while 
$Q^2_{max}\simeq s/2$.  Finally, the lower cut $Q^2_{min}$ is retained
as a free parameter which we will vary in order to optimise the
asymmetry $\Delta\s/\s$ while retaining a sufficiently high total
cross-section.

With these cuts, we find \cite{Narison:1992fd,Shore:2004cb}
\begin{equation}
\s \simeq 10^{-9} \frac{1}{Q^2_{min}} 
\log\frac{Q^2_{min}}{\L^2} \left(\log\frac{s}{Q^2_{min}}\right)^2 \ ,
\label{eg}
\end{equation}
while
\begin{equation}
\frac{\Delta\s}{\s} \simeq \frac{1}{2} \frac{Q^2_{min}}{s} 
\log\frac{s}{2 Q^2_{min}} \left[1 + \log\frac{s}{2\L^2}
\left(\log\frac{Q^2_{min}}{\L^2}\right)^{-1} \right] \ .
\label{eh}
\end{equation}
Notice in particular the relative $1/s$ supression of the polarization
asymmetry. This explains why a moderate energy $e^+ e^-$ collider is
best suited to the polarized QCD studies proposed here. The CM energy
of B factories such as Super-B, with $\sqrt{s}= 10.6~{\rm GeV}$, 
is ideal.

\begin{figure}[ht] 
\centerline{\includegraphics[width=2.8in]{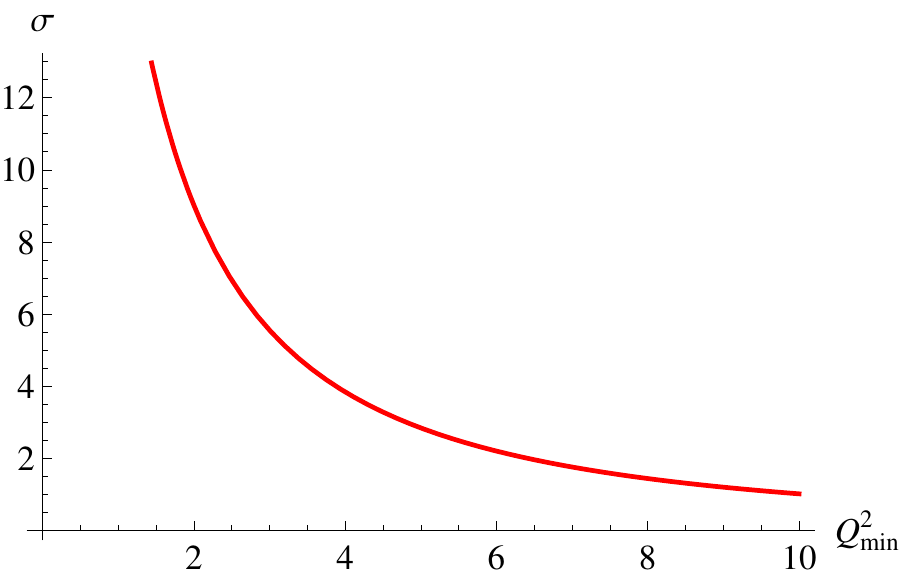}
\hskip0.7cm\includegraphics[width=2.8in]{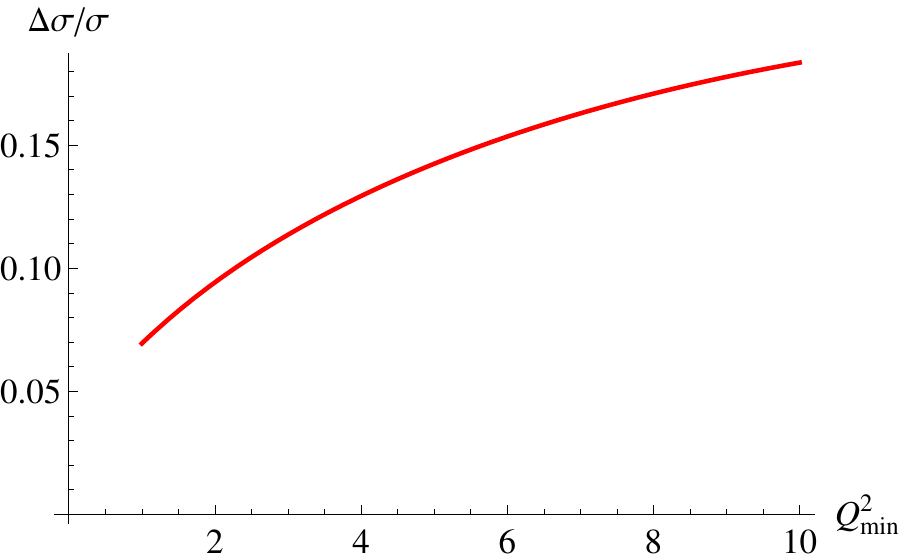}}
\caption{\footnotesize The total cross-section $\s$, measured in pb,  
for the inclusive reaction $e^+ e^- \rta e^+ e^- X$ plotted against
the lower cut-off $Q^2_{min}$ in ${\rm GeV}^2$.  The corresponding 
polarization asymmetry $\D\s/\s$ is shown in the right-hand figure.}
\label{Fig3}
\end{figure}

In Fig.~3, we plot $\s$ (in pb) and $\Delta\s/\s$ for the Super-B 
$\sqrt{s}$ over the range of $Q^2_{min}$ from 1 to 10 ${\rm GeV}^2$.
As is clear from the formulae above, we see that as $Q^2_{min}$ is
increased, the asymmetry $\Delta\s/\s$ increases, but at the 
cost of reducing the total cross-section $\s$. A reasonable compromise
is therefore to take $Q^2_{min} \simeq 2~{\rm GeV}^2$, which
gives $\s \simeq 10 ~{\rm pb}$ with an asymmetry 
$\Delta\s/\s = 0.1$.  

The extremely high luminosity of Super-B
means that this cross-section is sufficient to give a large
number of events. The design luminosity is $10^{36} ~{\rm cm}^{-2}
{\rm s}^{-1}$ and ref.\cite{Biagini:2010cc} quotes a potential annual
integrated luminosity $L = 12~{\rm ab}^{-1}$. 
With the above choice of the $Q^2_{min}$ cut, this corresponds to
$N = L\s = 10^8$ events/year, with a 10\% polarization asymmetry. 
To check the statistical significance of this asymmetry, we require 
$\D N/\sqrt{N} \gg 1$ and with these cuts we
find $\D N/\sqrt{N} = \sqrt{L\s} \D\s/\s \simeq 10^3$.
Of course, to measure the first moment sum rule for
$g_1^\c(x,Q^2;K^2)$ we need to distribute these events into sufficient
$Q^2$ and $K^2$ bins, but with such a high event rate even the
differential cross-section $d^2\D\s/dQ^2 dK^2$ should be easily
measurable with high precision.

The second main accelerator issue is polarization. In order to measure
the polarization asymmetry, both beams need to be polarized. At
present, the Super-B design only envisages polarizing the low-energy
beam, as required for example for $\tau$ polarization studies, 
but there appears to be no insurmountable technical 
obstacle to polarizing both beams given sufficient physics
motivation, which we believe an extensive programme of polarized 
QCD physics provides.

The polarization scheme designed for Super-B is described in detail 
in chapter 16 of ref.\cite{Biagini:2010cc} 
(see also \cite{Wienands:2010zz}). 
It involves the continuous injection of transversely
polarized electrons into the low-energy ring (LER) and subsequently 
use of an arrangement of spin rotator solenoids to bring the electron
polarization into longitudinal mode at the intersection region. 
The LER is chosen simply because the strength of the solenoids 
scales with energy. The design estimates that polarization
efficiencies in excess of 70\% at high luminosity can be sustained.

Finally, to measure the $K^2$ dependence of $g_1^\c(x,Q^2;K^2)$ in
detail in the dynamically interesting region $K^2 < 1.5~{\rm GeV}^2$,
we need to tag the `target' electron at sufficiently small angles,
since $K^2 = 4 E_2 E_2' \sin^2(\theta_2/2)$. Ideally, we would
like to be able to detect the electron at very small 
scattering angles $\theta_2$, to allow for small values of $K^2$ to be
measured for comparatively large energies $E_2'$.  
This raises the critical issue of detector acceptance.

The Super-B detector \cite{Grauges:2010}, which is based on a major
upgrade of BABAR, can detect particles with angles greater than
300 mrad to the beam direction
(see Fig.~1 of ref.\cite{Grauges:2010} for an overview sketch of the
planned detector).  
It is not clear whether it would be possible to add small angle
detectors capable of tagging an electron 
at angles around 50-100 mrad \cite{Kolanoski:1987sn} 
to the proposed design. However, as we now show, the comparatively 
modest beam energy of Super-B (taking $E_2 = 4.18~{\rm GeV}$ 
from the low-energy ring) means that even the detector acceptance 
of 300 mrad will in fact allow the target electron to be
tagged with the required values of $K^2$ while satisfying the
kinematical constraints for deep-inelastic scattering. 

\begin{figure}[ht] 
\centerline{\includegraphics[width=2.8in]{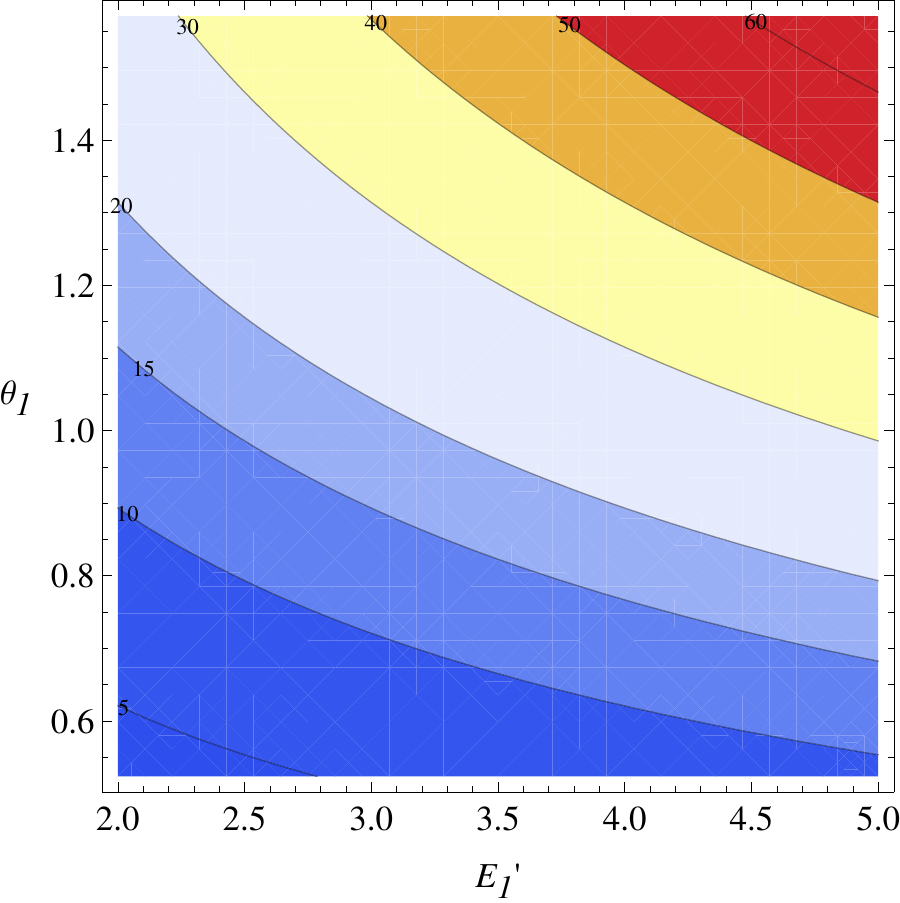}
\hskip0.7cm\includegraphics[width=2.8in]{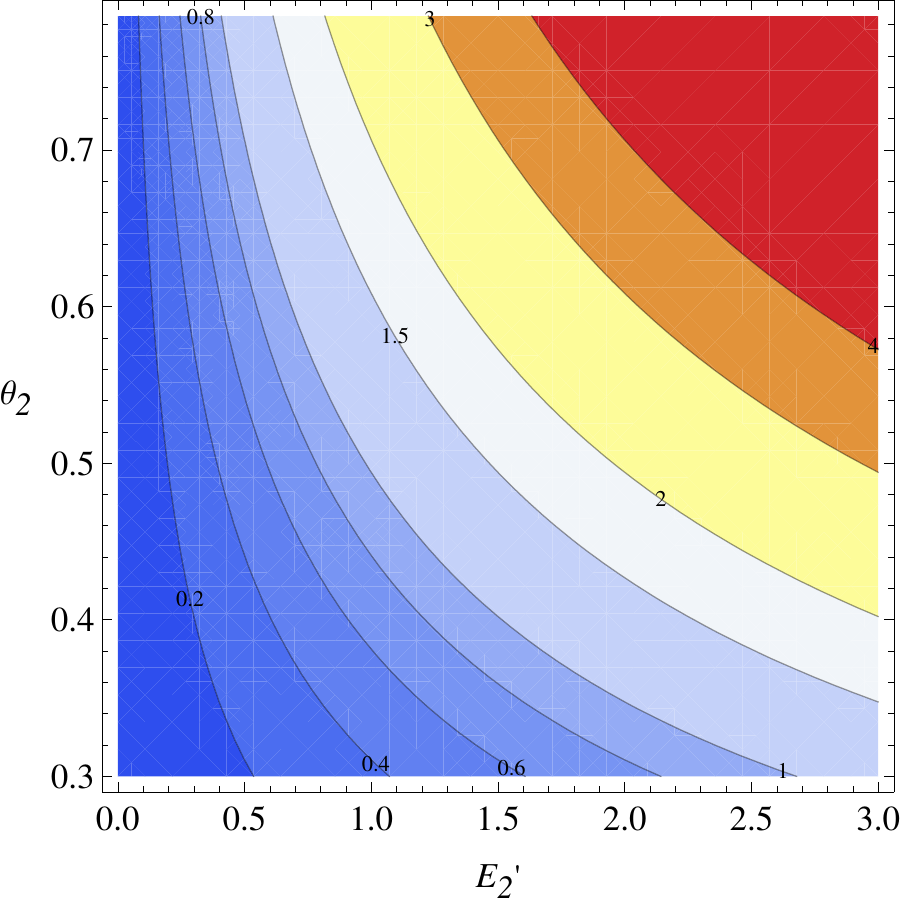}}
\caption{\footnotesize The left-hand figure (4a) shows a contour plot 
of $Q^2$ from 5 to 60 ${\rm GeV}^2$ for a range of 
electron scattering energies and angles $E_1'$ and $\theta_1'$.
The analogous contour plot for $K^2$ is shown in the right-hand 
figure (4b). This shows how the important region 
$0.2 < K^2 < 1.5~{\rm GeV}^2$ is accessible with target 
electron scattering energy in the range 
$0.5< E_2' < 2.5~{\rm GeV}^2$ with angle $\theta_2$ greater than
the detector acceptance 300 mrad. }
\label{Fig4}
\end{figure}

The dependence of $Q^2$ on $E_1'$ and $\theta_1'$, and the dependence
of $K^2$ on $E_2'$ and $\theta_2'$, are given in eq.\eqref{ba} and
illustrated in the contour plots in Fig.~4.  The $Q^2$ plot shows
that the full range of desired values from the optimal cut $Q^2_{min}
\simeq 2~{\rm GeV}^2$ up to $Q^2_{max}\simeq s/2$ can be easily
realised for scattering angles $\theta_1' > 300~{\rm mrad}$ and
energies $E_1'$ from 2-5 GeV. The required range 
$K^2 \lesssim 1.5~{\rm GeV}^2$ is shown by the series of curves in the
lower left of the contour plot Fig.~4b. This shows that the whole range
$0.2 < K^2 < 1.5~{\rm GeV}^2$ can be covered by tagging the electron
with $300~{\rm mrad}<\theta_2' < \pi/4$ and energy $E_2'$ in the range
$0.5 < E_2' < 2.5~{\rm GeV}$. We conclude that, coupled with the large 
number of events in this range guaranteed by the ultra-high 
luminosity, the Super-B detector will indeed be able to cover 
the range of $K^2$ and $Q^2$ necessary to measure the
$g_1^\c(x,Q^2;K^2)$ sum rule. 

\begin{figure}[ht] 
\centerline{\includegraphics[width=2.8in]{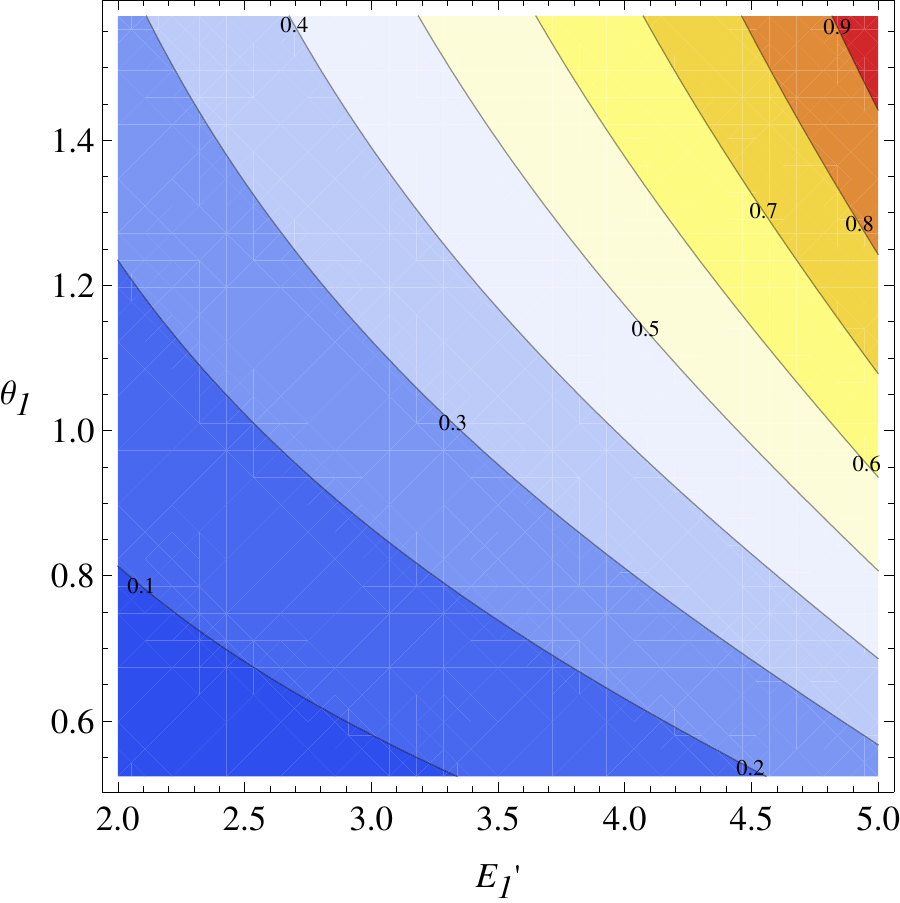}
\hskip0.7cm\includegraphics[width=2.8in]{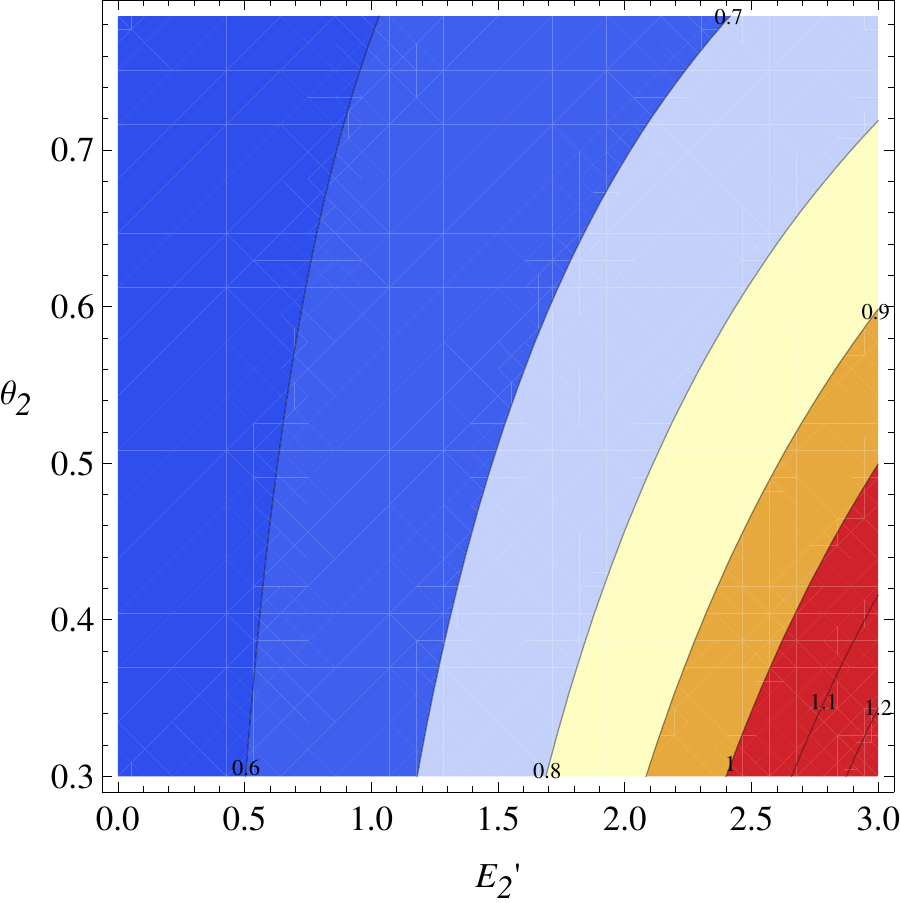}}
\caption{\footnotesize The left-hand figure (5a) shows a contour plot 
of $x_e = Q^2/2\n_e$ over a range of $E_1'$ and $\theta_1$.
The right-hand figure (5b) shows the corresponding plot for 
$x = Q^2/2\n$ as a function of $E_2'$ and $\theta_2$ for
values $E_1' = 4.5~{\rm GeV}$, $\theta_1 = \pi/3$.  
Satisfying the constraint $x < 1$ near the detector 
acceptance angle $\theta _2 \simeq 300~{\rm mrad}$ imposes
an upper bound $E_2' < 2.5~{\rm GeV}$. }
\label{Fig5}
\end{figure}

We also need to check the constraints on the remaining DIS variables.
From the derivation of the cross-sections involving $g_1^\c$, we
require the hierarchy $\n_e > \n > \tfrac{1}{2}(Q^2 + K^2)$, since
clearly the hadronic invariant momentum $W^2 > 0$,
as well as $x_e = Q^2/2\n_e < 1$ and $x = Q^2/2\n < 1$. 
Expressions for $\n_e$ and $\n$ in terms of the scattering energies 
and angles are given in eq.\eqref{ba}, and we also have the following
useful relation
\begin{equation}
\n_e - \n =  E_2'\Big[ E_1(1 + \cos\theta_2) - 
E_1'\big(1 + \cos(\theta_1 - \theta_2)
+ \sin\theta_1 \sin\theta_2(1 - \cos\phi)\big)\Big]
\label{ei}
\end{equation}
The condition $x_e < 1$ is satisfied provided only that $E_1'$ is not
too big, and from Fig.~5a we see that $E_1'\lesssim 5~{\rm GeV}$ is
sufficient. The condition $x<1$ is however much more stringent and
places an important upper bound on $E_2'$, which reduces as
the maximum value of $E_1'$ increases. From Fig.~5b we see that
to maintain $x<1$ with $E_2' \lesssim 2.5~{\rm GeV}$ at 
$\theta_2 = 300~{\rm mrad}$, we have 
to cut off $E_1'$ and $\theta_1'$ at around 4.5 GeV and $\pi/3$
respectively, corresponding to $Q^2 \simeq 30~{\rm GeV}^2$.
This is nevertheless a perfectly acceptable upper cut-off on
$Q^2$ for the DIS analysis. This maximum value of $E_2'$ is necessary
to optimise the available data in the required low $K^2$ region, as
shown above.  We can then show, using \eqref{ei} and numerical plots,
that $\n_e>\n$ provided $\theta_2$ is below an upper bound of
around $\pi/3$, so we confirm that the full hierarchy $\n_e > \n > 
\tfrac{1}{2}(Q^2 + K^2)$ holds for the range of variables already
determined.

In summary, all the kinematical constraints for the DIS analysis
of the first moment sum rule for $g_1^\c(x,Q^2;K^2)$ are satisfied 
with scattering angles and energies allowing measurements of 
$Q^2$ in the range $2 < Q^2 < 30~{\rm GeV}^2$ with $K^2$ 
satisfying $0.2 < K^2 < 1.5~{\rm GeV}^2$, provided we tag both
electrons with $2 \lesssim E_1' \lesssim 4.5~{\rm GeV}$,
$\theta_1 \lesssim \pi/3$ and $0.5 < E_2' < 2.5~{\rm GeV}$, 
$\theta_2 \lesssim \pi/4$ with both electrons scattered within
the detector acceptance $\theta_1,\theta_2 > 300~{\rm mrad}.$

\vskip0.1cm
Finally, we discuss briefly the prospects for measuring exclusive
processes at Super-B, such as the two-photon production of
pseudoscalar mesons through $e^+ e^- \rta e^+ e^- P$. 
The kinematics relating the invariants $Q^2,K^2,\n_e$ and $\bar\n_e$
to the observed electron scattering angles and energies is given in
eq.\eqref{ba}  as described above. In the exclusive case, however,
$\n$ is fixed by the constraint $W^2 = m_P^2$, while now the DIS
constraints on $Q^2$ and $\n_e$ are no longer relevant.
Tagging both electrons allows the off-shell meson transition functions
$g_{P\c^*\c^*}(Q^2,K^2)$ to be measured for essentially arbitrary
values of $Q^2$ and $K^2$, including very soft photons with $Q^2$
and/or $K^2$ down to around $0.2~{\rm GeV}^2$, limited only by the
detector acceptance. Knowledge of these transition functions will feed 
directly into eqs.\eqref{da} for the non-perturbative form factors
characterising the first moment of $g_1^\c(x,Q^2;K^2)$. 

The differential cross-section for $e^+e^-\rta e^+e^- P$ was derived
in eq.\eqref{di}. As already explained, since $\c^* \c^* \rta P$ for
pseudoscalar $P=\pi,\eta,\eta'$ is determined by a single form factor,
the transition functions can be obtained with unpolarized beams. The
polarization asymmetry would of course give new information for
two-photon production of higher-spin mesons which are characterised by
more than one form factor. Note also from \eqref{di} that in the
exclusive case, the polarization asymmetry of the differential
cross-section is suppressed by a double factor $O(Q^2K^2/s^2)$
compared to the single suppression $O(Q^2/s)$ for the
inclusive process. The ultra-high luminosity of Super-B
will allow a much-improved study of the off-shell transition functions
$g_{P\c^*\c^*}(Q^2,K^2)$ with both $Q^2$ and $K^2$ specified compared
to the existing data from CELLO \cite{Behrend:1990sr},
CLEO \cite{Gronberg:1997fj} and BABAR \cite{Aubert:2009mc,
BABAR:2011ad,Lees:2010de}.

\vskip0.1cm
In conclusion, we have shown how the unique combination of moderate
energy, polarization and ultra-high luminosity, together with its
detector capability, means Super-B has the ideal characteristics to
support an ambitious programme of two-photon QCD physics. This
includes, but is not limited to, the investigation of pseudoscalar
meson transition functions, with their relevance to the muon $g-2$, 
and the photon structure functions 
$g_1^\c(x,Q^2;K^2)$ and $g_2^\c(x,Q^2;K^2)$, including the potential
to make the first experimental measurement of the first moment sum
rule for $g_1(x,Q^2;K^2)$. This will give direct experimental input
into many interesting theoretical issues in QCD, including chiral
symmetry breaking, $U(1)_A$ dynamics, gluon topology and anomalous
chiral symmetry. All this provides strong motivation for including
polarized two-photon QCD physics as an important element of the 
research programme planned for Super-B.

\vskip0.7cm
\centerline{*******}

\noindent I would like to thank S.~Narison and G.~Veneziano for their
original collaboration on the photon sum rule and the Theory
Division, CERN for hospitality during the course of this work. 
I am grateful to the U.K.~Science and Technology Facilities Council
(STFC) for financial support under grant ST/J000043/1.

\end{document}